\documentclass[12pt,preprint]{aastex}                                    

\usepackage{rotating,lscape,longtable}

\newcommand{\htwo}{H$_2$}
\newcommand{\ctwo}{C$_2$}
\newcommand{\cthree}{C$_3$}
\newcommand{\ebv}{$E_{B-V}$}
\newcommand{\HI}{H$\;${\small\rm I}\relax}
\newcommand{\NaI}{Na$\;${\small\rm I}\relax}
\newcommand{\KI}{K$\;${\small\rm I}\relax}
\newcommand{\NHI}{$N({\rm H \; \mbox{\small\rm I}})$}
\newcommand{\NH}{$N$(H)}
\newcommand{\NCH}{$N$(CH)}
\newcommand{\Nhtwo}{$N$(H$_2$)}
\newcommand{\NHtwo}{$N$(H$_2$)}

\newcommand{\NNa}{$N$(Na)}
\newcommand{\kms}{km~s$^{-1}$\relax}
\newcommand{\etal}{et al.}
\newcommand{\lam}{$\lambda$}

\newcommand{\chisq}{$\chi^2$}

\newcommand{\Wlam}{$W_{\lambda}$}

\begin{document}

\title{Studies of Diffuse Interstellar Bands. V. Pairwise Correlations of Eight Strong DIBs and Neutral Hydrogen,
Molecular Hydrogen, and Color Excess}

\author{Scott D. Friedman\altaffilmark{1},
Donald  G.  York\altaffilmark{2},
Benjamin J. McCall\altaffilmark{3},
Julie Dahlstrom\altaffilmark{4},
Paule Sonnentrucker\altaffilmark{5,1},
Daniel E. Welty\altaffilmark{6},
Meredith M. Drosback\altaffilmark{7},
L.  M.  Hobbs\altaffilmark{2,8},
Brian  L.  Rachford\altaffilmark{9},
Theodore  P.  Snow\altaffilmark{10}}

\altaffiltext{1}{Space Telescope Science Institute, Baltimore, MD 21218; friedman@stsci.edu}

\altaffiltext{2}{Department of Astronomy and Astrophysics and the Enrico Fermi Institute, University of Chicago}

\altaffiltext{3}{Departments of Chemistry, Astronomy, and Physics, University of Illinois at Urbana-Champaign}

\altaffiltext{4}{Department of Physics and Astronomy, Carthage College}

\altaffiltext{5}{Department of Physics and Astronomy, Johns Hopkins University}

\altaffiltext{6}{Department of Astronomy, University of Illinois at Urbana-Champaign}

\altaffiltext{7}{Department of Astronomy, University of Virginia}

\altaffiltext{8}{Yerkes Observatory, University of Chicago}

\altaffiltext{9}{Department of Physics, Embry-Riddle Aeronautical University}

\altaffiltext{9}{Center for Astrophysics and Space Astronomy, University of Colorado}

\begin{abstract}
We establish correlations between equivalent widths of eight
diffuse interstellar bands (DIBs), and examine their correlations
with atomic hydrogen, molecular hydrogen, and \ebv. The DIBs are centered at
$\lambda\lambda$ 5780.5,  6204.5, 6283.8, 6196.0, 6613.6,
5705.1, 5797.1, and 5487.7, in decreasing order of Pearson's
correlation coefficient with \NH\ (here defined as the column density of neutral hydrogen),
ranging from 0.96 to 0.82.
We find the equivalent width of $\lambda$5780.5 is better
correlated with column densities of H than with \ebv\ or \htwo, confirming earlier
results based on smaller datasets.  We show the same is
true for six of the seven other DIBs presented here.  Despite this similarity, the
eight strong DIBs chosen are not well enough correlated with each other to
suggest they come from the same carrier.
We further conclude that these eight DIBs are more likely to be associated
with H than with \htwo,  and hence are not preferentially located in the densest,
most UV shielded parts of interstellar clouds. We
suggest they arise from different molecules found
in diffuse H regions with very little \htwo\ (molecular fraction f$<$0.01).
Of the 133 stars with available data in our study, there are three
with significantly weaker $\lambda$5780.5
than our mean H -- $\lambda$5780.5 relationship, all of which are in regions of
high radiation fields, as previously noted by Herbig. The correlations will be
useful in deriving interstellar parameters when direct methods are not available.
For instance, with care, the value of \NH\ can be derived from W$_{\lambda}(5780.5).$

\end{abstract}


\section{Introduction}

The diffuse interstellar bands (DIBs) represent a long standing, spectroscopic mystery: hundreds of weak
absorption features detected in the optical wavelength range remain
unidentified (see Herbig 1995; Snow 1995, 2001 for summaries). While they were noted in stellar
spectra as early as 1919 (Heger 1922), the DIBs had their suspected interstellar nature demonstrated
more than a decade later (Merrill 1936). \\

Most of the early hypotheses regarding the progenitors (carriers) of DIBs centered on molecules, but
by the early 1970s solid-state (i.e., grain) carriers were thought to be more likely (Herbig 1975).  
Molecules were re-introduced in the mid-1970s (Danks and Lambert 1976; 
Douglas 1977; Smith, Snow, and York 1977), and now most researchers have adopted 
large molecules or their ions as the most likely candidates (see Herbig 1995 for a review).
Efforts to match laboratory spectra with observed DIB profiles have not been successful.
Tulej et al. (1998) reported a match between the laboratory spectrum of C$_{7}^{-}$ and five
narrow DIBs, but with improved laboratory and astronomical data this was subsequently
shown to be incorrect (McCall et al.
2001). Motylewski et al. (2000) found a weak astronomical feature at approximately the same
laboratory wavelength and profile as HC$_5$N$^+$. More recently, Krelowski et al. (2010)
have suggested that the laboratory spectrum of HC$_4$H$^+$ closely matches a newly
identified, weak DIB at 5068.8\,\AA. However, confirmation of these claims by a match with
a second line in laboratory and astrophysical spectra, has not yet been made.
Thus, the carriers still remain unidentified. \\ 

The DIBs have, thus, become recognized as a new window into the chemistry of the interstellar
medium -- if we could only identify their carriers.  Attempts to identify the DIBs have included (i)
searches for molecules in the laboratory with the same spectroscopic signatures  as the DIBs
(Leach 1995; Herbig 1995; Allain et al. 1996; Salama et al. 1996, 1999; McCall et al. 2000);
(ii) modeling of the structures detected in some DIB profiles in terms of rotational excitation of
gas-phase molecules (Cossart-Magos \& Leach 1990; Sarre et al. 1995; Galazutdinov et al. 2002b,
2008); and (iii) searches for correlations of DIBs with other interstellar  parameters (Wampler 1963, 1966;
Snow, York and Welty 1977; Sneden et al. 1978; Wu, York, and Snow 1981; Herbig 1993;
Jenniskens and D\'esert 1994; Sonnentrucker et al. 1997, 1999; Thorburn, et al.  2003;
Weselak et al. 2004, 2008).\\

Finally, searches for correlations between individual pairs of DIBs were also carried out for the
purpose of finding whether some of the DIBs were better correlated with each other than other pairs of 
DIBs.  The reasoning was that these studies could reveal sets of DIBs that came from the 
same or similar carriers. The works of Krelowski and Walker (1987); Josafatsson and Snow (1987);
Westerlund and Krelowski (1989), Cami et al. (1997), and Weselak et al. (2001) led to the
identification of ``families'' of
DIBs.  The DIB pair thought to show the best correlation is comprised of the $\lambda$6196.0 and
$\lambda$6613.6\footnote{Due to uncertainty in the rest wavelength of DIBs, and differing
practices by various authors for truncating or rounding wavelengths, the nomenclature
of DIBs in the literature is confusing. Indeed, with the increasing number of known DIBs
(Hobbs et al. 2008, 2009) quoting wavelengths to only integer values can be ambiguous.
Therefore, we use the wavelengths that are tabulated in Table 2 of the Hobbs et al. (2008)
study of HD 204827, and
round to five significant figures. For the three narrow DIBs, \lam\lam5780.5, 6613.6, and 5797.1,
the central wavelength found by Hobbs et al. (2009) for HD 183143 are 0.1\,\AA\ longward of the
wavelengths given here. This may be the result of component structure differences in the velocity
profiles for these sight lines.}  DIBs (Cami et al. 1997; Moutou et al. 1999; Galazutdinov et al. 2002b).
Since no observed correlation was perfect, agreement on which DIB belonged to
which family, or whether pairs of DIBs arise from the same carrier, was not always
reached when comparing these studies.  \\

To address these issues we compiled a large spectral database toward approximately 200 stars,
which has generated a series of papers on diverse properties of DIBs. Thorburn et al. 2003 (Paper I)
describe the relationship between C$_2$ and certain DIBs. Hobbs et al. 2008, 2009 (Papers II and III)
present spectral atlases of DIBs toward the spectroscopic binary star HD 204827 and toward HD 183143.
McCall et al. 2010 (Paper IV) revisited the $\lambda\lambda 6196.0, 6613.6$ correlation.
The unprecedented data quality and statistics of our survey (see \S2) show that this pair has the
highest correlation of any known pair and the data would be consistent with a perfect correlation
if the errors were underestimated by only a modest amount. In the present paper, we extend
our investigation to more fully study the eight strong DIBs $\lambda\lambda$ 5780.5, 6204.5,
6283.8, 6196.0, 6613.6, 5705.1, 5797.1, and 5487.7, in order of decreasing
Pearson's correlation coefficient with \NH\footnote{We
represent the column density of neutral atomic hydrogen by \NH. This is often mistakenly
denoted \NHI. However, \HI\ actually denotes the spectral line of atomic hydrogen.}.
We also examine the DIB-DIB correlations as well as the correlation of
the DIBs with the column density of molecular hydrogen and with color excess. \\

This paper is organized as follows. In \S2, we briefly describe how the survey data were 
obtained and reduced, and we present an extensive list of the line of sight parameters and DIB
equivalent widths (EWs) toward the 133 stars reported here. In \S3 we present a large variety of
correlation coefficients and plots, and the slopes and intercepts of correlation plots between
$\lambda5780.5$ with the other DIBs and with \NH, \Nhtwo, and \ebv. In \S4 we discuss
these results, including the importance of systematic errors which arise among measurements
of DIB equivalent widths. In \S5 we summarize the results of this study.\\

\section{Observations and Data Analysis}

From 1999 to 2002 we obtained a high signal-to-noise data set on DIBs in the spectra
of about 200 stars spanning a large range of reddening, \ebv\ $\sim$ 0.01 to 3.31 magnitudes
(Papers I, II, and III) and their associated diatomic or triatomic molecules (Oka et al. 2003). The
reader is referred to those papers for details, and we give only a brief description of the
data analysis here. The echelle spectrograph (Wang et al. 2003) was used on the Apache
Point Observatory 3.5-meter telescope to obtain spectra at a resolving power
$\lambda$/$\Delta\lambda = 38,000$ from
3600\,\AA\ to 9000\,\AA\ at a nominal signal to noise ratio of roughly 1000 at 5780 \AA\ for each
sight line (see Paper I). Stellar lines are distinguished from DIBs in reddened stars by comparison
with stars of the same spectral type but with low reddening. Telluric lines are
removed by use of a complex scheme that measures patterns of behavior
in key telluric absorption lines and makes a blanket correction for each observation,
depending on air mass and humidity. DIBs that were generally uncontaminated by stellar
and telluric blends were measured as described extensively in Papers I and II.

DIB absorption features pose special challenges for any study, such as this one, seeking
to quantify equivalent widths. Ultimately, the goal of DIB equivalent width measurements
must be to include all blended absorption from the same chemical species without
including blended absorption from other chemical species. However, correctly
distinguishing contaminating features from those belonging to the same compound
itself presupposes knowledge of those chemical species and their spectra.
As long as the DIBs carriers remain unidentified, the shape and width of the spectral
profiles will remain uncertain. The band structure could be due either to blending
with features from other carriers or to blending with features from
higher rotational levels of the same species, or could be due to both types of blends.

For this study, and for the previous papers in this series, continuum normalization was
accomplished by use of eighth-order Legendre polynomials. For the broadest DIBs,
absorption spanned multiple spectral orders. In these cases continuum profiles were
estimated by interpolation across multiple orders. Equivalent widths were hand-measured
by one of us (JD), along with inspection of each DIB in comparison with nearly
unreddened standard stars in order to identify stellar lines.
Limits of integration were set by where the DIB absorption recovered to the continuum.
Where the profile did not fully recover to the continuum, we set the limits at inflection
points in the profile, which often indicate an appropriate endpoint (Krelowski \& Sneden,
1993). We made no assumption that the line shapes have Gaussian profiles.
Equivalent widths were computed with direct integration.

We do not presume that the present choices of integration endpoints are in any way
definitive, nor do we suppose that any prior work has made similar choices with any
more assurance of correctness or fewer reasonable arguments in favor of their choices.
Integration endpoints chosen in the present study reflect a primary interest in
\emph{repeatable} measurements and minimal sensitivity to continuum location error,
both of which are essential to reducing the scatter in correlations from measurement
error but do not prevent scatter from other, less easily avoided causes. We do not
recommend that the present equivalent widths be combined from those reported in
studies by other investigators without carefully considering the specific definitions of
what constitutes the measurement criteria for each diffuse band.

The quantity used to investigate correlations differs from study to study, even among
the same investigators. For example, Moutou et al. (1999) used central depths, stating
that they are less sensitive to contamination than are EWs, although they point out
that depths reflect well the correlation in EW. Weselak et al. (2001) find in their study
that central depths correlate better than EWs. On the other
hand, in a study at very high spectral resolution (220,000) of the profiles of
\lam\lam6196, 6614, Galazutdinov et al. (2002a) found a correlation in the
equivalent widths of the DIBs but not in the FWHMs. The FWHM of \lam6196 varies
by 50\% over the 7 sight lines sampled, whereas for \lam6614 it is nearly
constant within observational errors. The substructure of both profiles varies among
the sight lines but in an unrelated way for the two DIBs. Interstellar
atomic lines do not reveal Doppler broadening, so the authors believe the absorption
arises in a single cloud. 

Until the identification of DIBs is secure, by definitive matches with laboratory spectra,
no single method can be deemed superior. We have elected to base our
correlation studies on measurements
of equivalent widths rather than FWHM or the central depths of profiles. While the
latter two measures are less likely to suffer from contamination by nearby, unrelated
species, they will fail to account for broad absorption due to R or P branch transitions, for
example. The strength of these branches depends on the unknown rotation temperature
of the molecules along the line of sight, which can vary from cloud to cloud, and even
within a cloud. As shown
in Figure 1 of Oka at al. (2003), the absorption can occur over a rather large wavelength
interval. Another reason to use EWs rather than central depth is that most of these sight
lines intersect multiple clouds, so our results represent averages over the intervening clouds.
Doppler splitting could affect measures of central depth, especially for narrow DIBs.
As long as there is not saturated component structure, EWs will not suffer from
this error. Finally, measures of equivalent widths are independent of the resolution of the
instrument used, which is not true of FWHM or central depth.

Standard techniques were used for cosmic ray
removal, flat fielding, background and bias subtraction, and
extraction to one dimensional spectra. The
spectra from adjacent orders were coadded to give a blazeless
spectrum. A set of 35 lines that were most free of stellar
blending or telluric line contamination were initially measured.
The estimates of correlations among DIBs and between DIBs and other
interstellar quantities are based on DIB measurements presented by
Paper I for an expanded set of stars, on the H and \htwo\ column densities of
Rachford et al. (2002, 2009), and on color excess values collected from the
literature by one of us (LMH) based on the color scale of
Johnson (1963). Errors on each data point were estimated by measurements in adjacent
parts of the continuum, free of other DIBs
and of stellar or telluric lines, and then were propagated through the analysis.
One sigma errors are used throughout. McCall et al. (Paper IV) give an extensive
discussion of additional errors that may affect the data. They suggest
that our errors may be underestimated by about a factor of 2 due to
systematic effects, the three most likely of which are continuum placement
errors, the possible presence of unidentified weak DIBs close to the
DIBs being measured, and the residual errors arising from imperfect
removal of telluric water vapor lines. In the present study, however, we use the
formally propagated errors described in Paper I.

Table 1 includes a list of all 133 stars in this study, their spectral types and luminosity class,
\NH, \NHtwo, \ebv, and the equivalent widths of the eight DIBs we focus on here.
This analysis excludes the 17 lines noted as being
correlated with \ctwo\ (Paper I) and \cthree\ (Oka et al, 2003), whose relationships with \ion{K}{1}, CO, CH,
and other molecular species will be the subject of a future study by our group.
Rachford et al. (2009) did a reanalysis of \NH\ and \NHtwo\ along some of the sight
lines included in the current study. The values that they found of these quantities differ
from those derived in the original analysis by less than $1\sigma$ in all cases, much less
than the cosmic dispersion of these quantities. We therefore use the results of our original,
homogeneous data analysis, as presented by Rachford et al. (2002).

\section{DIB Profiles and Correlation Properties}

The DIBs chosen for this study were meant to avoid the broadest DIBs, for which our echelle
measurements can underestimate the line strengths (Hobbs et al. 2009) due to continuum
placement difficulties. They were also meant to focus on the classic strong DIBs, which do
not include the C$_2$ DIBs found by Thorburn et al. (2003) (Paper I). The strongest DIBs
with FWHM less than that of \lam 5780.5 are
included (\lam\lam5797.1, 6613.6). The DIB at \lam6283.8 is the second strongest DIB
after \lam4428.1, so it was included to test for any evidence of saturation in the range of
\ebv\ for our stars, even though it has FWHM = 4.77\,\AA\ (Table 2). Another much weaker
DIB, \lam5487.7, was included as a comparison. Two lines were
included because of their previously known correlations with other DIBs: \lam 5705.1
(used in Paper I to test for saturation in \lam 5780.5) and \lam 6196.0,
previously noted for its close correlation with \lam6613.6. An eighth line (\lam 6204.5)
was included since it has two narrow components that are blended and it is not clear if
they should be measured together or separately. This characteristic is shared by other
DIBs, such as \lam5849.8 and \lam6660.7, but we selected \lam6204.5 because it is among
the strongest DIBs. The DIBs in Paper I were analyzed with respect to their
correlation with each other. The DIBs chosen for this paper are at the high end of correlation
coefficients and include all the moderately well correlated DIBs. All other DIBs
from Paper I are less well correlated with each other than those in this paper.

\begin{figure}
\epsscale{0.80}
\plotone{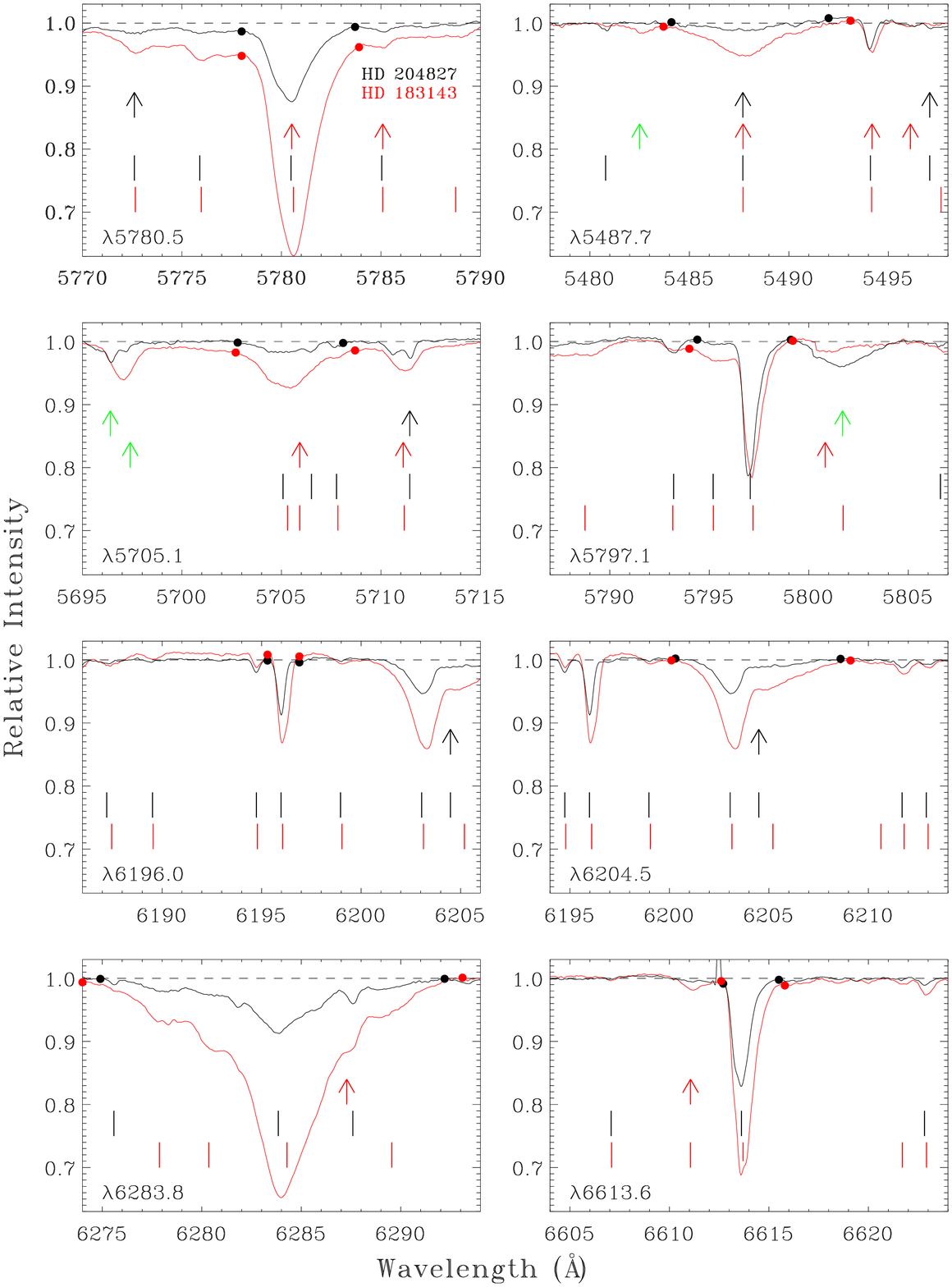}
\caption{Continuum normalized spectral profiles of the eight DIBs toward HD 204827 (black)
and HD 183143 (red). Filled circles indicate the limits of integration for calculating equivalent widths.
Black arrows indicate the locations of stellar lines
identified in the DIB atlas for HD 204827 (Paper II) and red arrows the stellar lines for HD 183143
(Paper III). Green arrows show additional stellar lines identified in the low-reddened comparison
stars for these two stars. Note the apparent offset in LSR velocity of the DIBs in the spectra of the
two stars. See text for an explanation. The vertical scale in all panels is the same to clearly show
the relative strengths of the DIBs. The spike just to the left of the $\lambda6613.6$ DIB is an artifact.
\label{dib8panel}}
\end{figure}

Spectral profiles of these eight DIBs toward HD 183143 and
HD 204827 are shown in Figure~\ref{dib8panel}. HD 183143 has been observed by
many DIB investigators (e.g., Herbig 1975; Jenniskens \& Desert 1994; Tuairisg et al. 2000)
and is the basis for the atlas discussed in Paper III. HD 204827, discussed in Paper II, reveals
several narrow, weak DIBs which are not evident in HD 183143, and vice versa. Some broad
DIBs are not seen in common in the two stars, as well. 
We note that the central wavelength for each band in Fig.~\ref{dib8panel} is slightly longer for
HD 183143 than for HD 204827, a systematic difference anticipated and discussed in some detail
in Papers II and III. The unavoidable uncertainty in the precise zero-point for DIB wavelengths
arises primarily from the combination of the multiple interstellar clouds present along both stellar
lines of sight and the unknown identities and laboratory spectra of the molecules presumed to
cause the DIBs. The median offset for the eight bands in Fig.~\ref{dib8panel} corresponds to
about 7 km/s. Had we arbitrarily chosen to assign the laboratory \KI\ wavelength to the mid-point
between the two main components of the interstellar \KI\ line toward HD 183143 (Paper III, Fig. 2),
for example, this systematic median offset would effectively be removed. This result implies that
the DIBs show component structure. Indeed, Doppler splitting in the narrow DIB \lam6196.0, matching
the splitting of interstellar \KI, about 15 \kms, was observed in high-resolution spectra of HD 183143
by Herbig \& Soderblom (1982). However, in our observations the scatter in the mean is such
that some DIBs may be dominant in one component and some in others. The offsets for the
narrowest DIBs are clear. We will address this issue in a future paper (York et al., in preparation).

Figure~\ref{dib8panel} also illustrates our choice of integration limits for the DIBs along these
two sight lines, and the ambiguities this involves. For example,  \lam5780.5 is either at the
bottom of a broad feature or else is flanked by a series of narrow, possibly related, features.
Not knowing the origin of each feature, we assumed the former interpretation as did, for
example, Galazutdinov et al. (2004). For \lam6204.5 we include the extended red wing.
Porceddu, Benvenuti, \& Krelowski
(1991) conclude that \lam6205 is a separate DIB on the basis of a varying central depth ratio
of the two features. They further conclude that these differences are due to differences
in the physical parameters within a single interstellar cloud, based on the lack of observed
Doppler splitting in the narrow \lam6196.0 DIB, at their moderate resolving power (30,000)
and SNR (200--300). For \lam5797.1 we include the blue wing, in contrast to Galazutdinov
et al. (2004), who believe it is blended with a much broader feature at \lam5795. As shown in
\S3.2, we find very good correlations associated with \lam6204.5 and the lower ones for
\lam5797.1. Determining which features correlate best may prove to be a useful tool for
guiding the placement of integration
limits. However, it is impossible to know the correct approach in advance, so we favor
systematic repeatability until the true profile can be established by species identification
in the laboratory.

Also shown in Figure~\ref{dib8panel} are the DIBs identified in atlas Papers II and III (black
and red tick marks), the stellar lines (black and red arrows), and a few additional stellar lines
(green arrows) observed in the spectra of the lightly reddened comparison stars. Broad DIBs,
like \lam6283.8, may include multiple weak DIBs identified in the atlases. The criteria for selecting
the DIBs is clearly described in these papers, but no claim is made regarding whether
these separate DIBs really arise from the same carrier molecule. Thus, including or excluding
them are equally justifiable approaches. Note that our choice to include them does not
significantly affect our results. For example, the 3 DIBs in the wings of the HD 183143
profile have equivalent widths of 28, 22, and 19 m\AA\ (Paper III) which, even in total
comprise only a few percent of the 1910 m\AA\ equivalent width of the main DIB. Similarly, for
HD 204827 the EWs of the 2 flanking DIBs are 3 and 14 m\AA\ (Paper II), compared to
518 m\AA\ for the main DIB. For \lam5797.1 and \lam6204.5, flanking DIBs were formally
identified in the atlas papers but, as noted above, we elected to include the wings in our
EW measurements of the main DIBs until identifications are secure.

There are very few stellar lines blended with the DIBs considered here. Those present are generally
weak. Their treatment with respect to measuring equivalent widths is described in Papers II and III.

\subsection{Correlation of $\lambda 5780.5$ with \NH, \NHtwo, and \ebv}

\begin{figure}
\epsscale{1.0}
\plotone{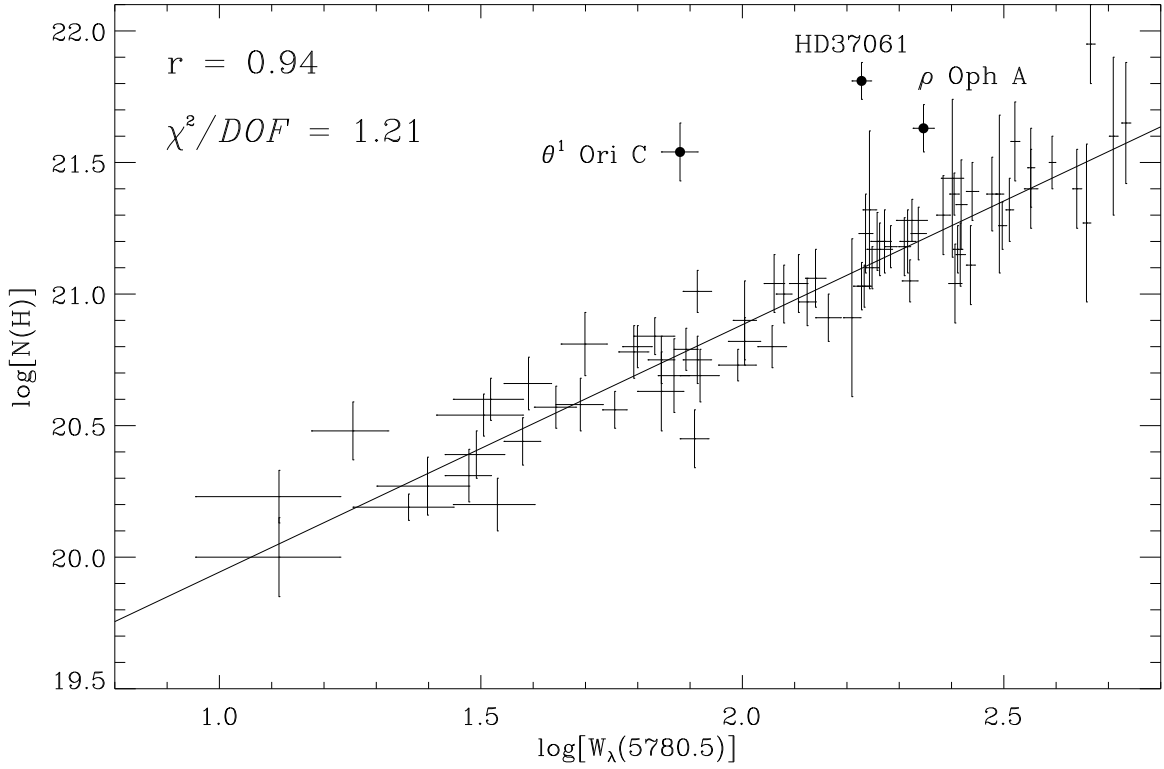}
\caption{Log[\NH] $vs.$ log[$W_{\lambda}(5780.5)$].  In this and the figures which follow,
the straight line is the least squares fit to the data excluding the
outlying points, which are indicated by filled circles. The slope and intercept
of the line are given in Table 2. All equivalent widths are in units of m\AA.
\label{nh_5780}}
\end{figure}

Herbig (1993) noted the strong correlation between \NH\ and W$_{\lambda}(5780.5)$.
This relationship is shown in Figure~\ref{nh_5780} for our sight
lines. The correlation coefficient\footnote{All correlation coefficients
in this paper refer to Pearson's correlation coefficient. The reader is cautioned that some
authors (e.g. Wallerstein, Sandstrom, \& Gredel 2007) use other statistics, such as Spearman's
rank correlation coefficient.} for this relationship is $r = 0.94$, when 3 stars are excluded
from the sample: $\rho$ Oph A, $\theta^1$ Ori C, and HD 37061 (see Table 2). These
outliers were also rejected by Herbig (1993), who
noted the remarkably high radiation field of the Trapezium stars, and that both
\NNa\ and the DIBs $\lambda$5780.5 and $\lambda$5797.1 are low with respect to \NH.
These three stars are very weak in \ion{K}{1} (Welty \& Hobbs 2001). They have very flat
far-UV extinction curves and $\theta^1$ Ori C and HD 37061, in particular, have weak
2175\,\AA\ bumps (Fitzpatrik \& Massa 2007).
Table 2 also includes the reduced \chisq\ for a straight line fit to the plot
of log($\lambda5780.5$) $vs.$ log(\NH), excluding these three outlying stars.
The values in the fifth column of the table are the correlation
coefficients when these stars are included in the sample. Table 2 also includes
the correlation coefficients for the remaining DIBs with log(\NH), as well as the
coefficients of the best linear fits, excluding the same three outliers.

\begin{figure}
\epsscale{1.0}
\plotone{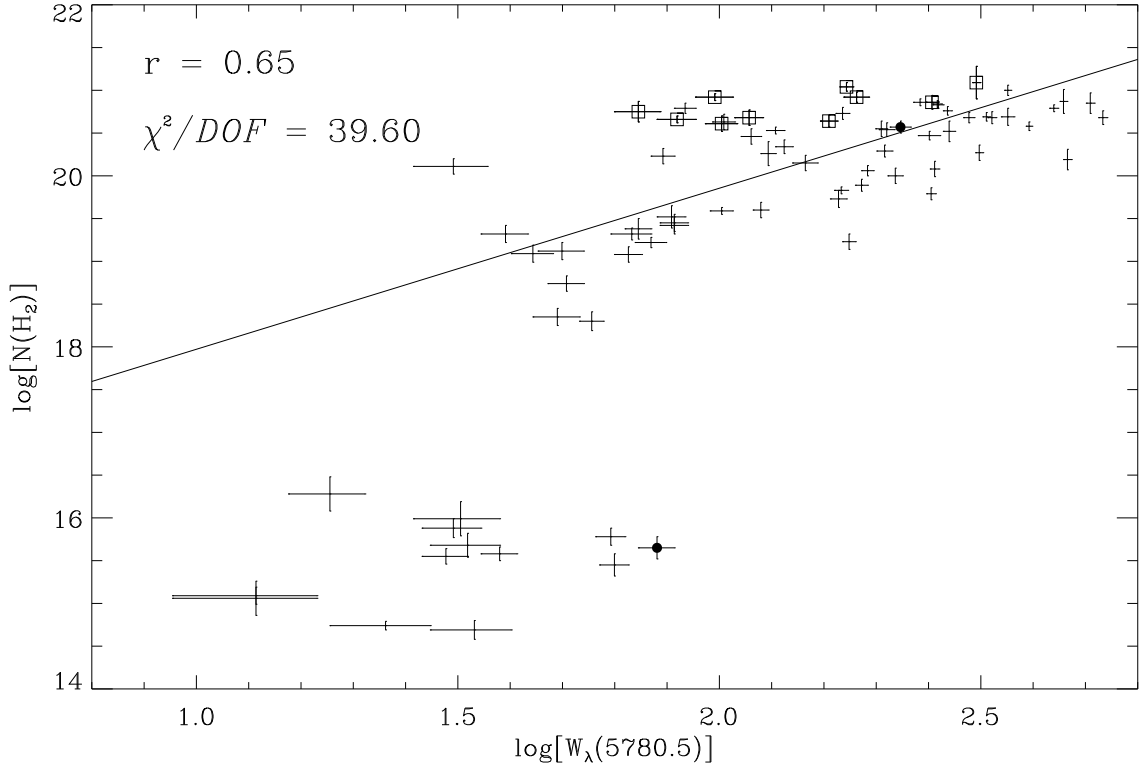}
\caption{\Nhtwo\ $vs.\; W_{\lambda}(5780.5)$. Open squares
denote sight lines with molecular fraction $f$(\htwo) $>$ 0.5.
The straight line is the least squares fit to the points with
log[\Nhtwo] $> 18$, and is given by log[\Nhtwo] = $(16.09 \pm 0.09)$ +
$(1.88 \pm 0.04)$ $\times$ log[$W_{\lambda}(5780.5)$].
\label{nh2_5780}}
\end{figure}

Figure~\ref{nh2_5780} shows the relation between \Nhtwo\ and $W_\lambda(5780.5)$.
For the full data set it is clear that there is a very
poor relationship between the two quantities, especially compared to
Figure~\ref{nh_5780} for \NH\ and $W_{\lambda}(5780.5)$. It is noteworthy
that stars with \Nhtwo\ ranging over a factor of more than $10^5$ can have
identical values of $W_{\lambda}(5780.5)$.
A special set of sight lines is denoted by open squares in Figure~\ref{nh2_5780}:  those with
fractional abundance of \htwo\ ($f$ = [2\Nhtwo]/[\NH + 2\Nhtwo]) greater than 0.5.
This molecular fraction is a line-of-sight average, so the value of $f$ for some
individual clouds almost certainly must be greater than 0.5. In these clouds most hydrogen
is in the form of \htwo. Considering a slightly less restricted set of sight lines, those
for which log(\Nhtwo)$>18$ we find $r = 0.65$, considerably lower than the correlation
coefficient of \NH\ with $\lambda5780.5$. The correlation coefficients
for all eight DIBs with \Nhtwo\ are given in Table 3. The reduced $\chi^2$ values are far
greater than for the DIB--log(\NH) relations shown in Table 2.

\begin{figure}
\epsscale{1.0}
\plotone{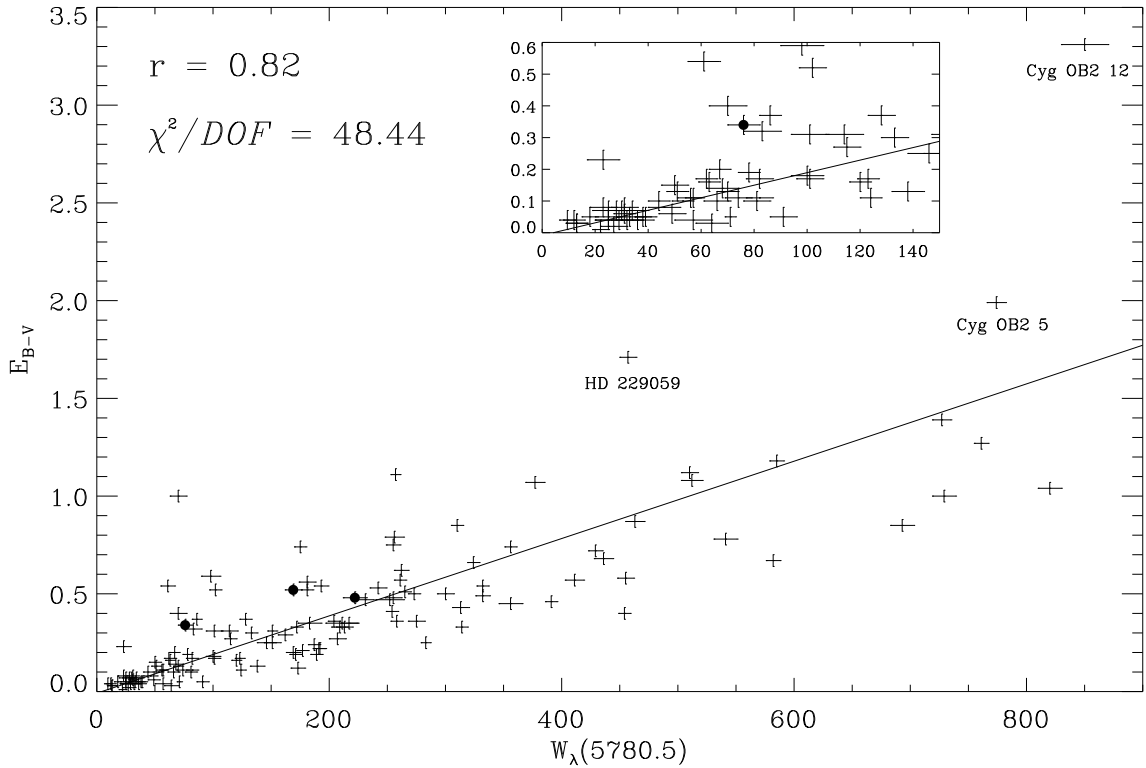}
\caption{\ebv\ $vs.$ $W_{\lambda}(5780.5)$. The inset in this and the following
correlation plots shows a close-up view of the region
near the origin. The best fit line has not been
constrained to go through the origin in any of the plots in this paper.
\label{ebv_d5780}}
\end{figure}

Figure~\ref{ebv_d5780} shows the correlation of \ebv\ and $W_{\lambda}(5780.5).$
Reddening errors are difficult to quantify, but we estimate them to be approximately 0.03 mag.
This plot covers a large range in \ebv, from 0.01 to 3.31 mag. The three points with the
highest values of \ebv\ are Cyg OB2 12, Cyg OB2 5, and HD 229059, all of which are
in the Cyg OB2 cloud.There is significant scatter in this plot, with $r=0.82,$ indicating that
$\lambda5780.5$ is not directly associated with the column density of dust grains responsible
for the optical differential extinction. The correlation coefficients
for all eight DIBs with \ebv, as well as the coefficients for the best fit lines, are given in
Table 4. Again, the reduced $\chi^2$ values are far
greater than for the DIB--log(\NH) relations. Note that the best fit lines have not been
constrained to go through the origin in any of the plots presented in this paper.

The prominent jump in the data points in Figure~\ref{nh2_5780} is at $W_\lambda(5780.5)
\approx$ 50m\AA, corresponding to
\ebv $\approx 0.1$, according to Figure~\ref{ebv_d5780}. This matches well the value of
\ebv $\approx 0.08$ that marks the beginning of a sharp transition from low to high
values of \Nhtwo\ in interstellar clouds (Savage et al. 1977). This level of
reddening indicates the presence of a sufficient density of dust grains to favor the
formation of \htwo, and a high enough column density of \htwo\ initiates self-shielding, which
allows an exponential increase in \Nhtwo.

\subsection{DIB-DIB Correlations}

The eight DIBs in our sample are
$\lambda\lambda$ 5780.5, 6204.5, 6283.8, 6196.0, 6613.6, 5705.1, 5797.1, and
5487.7, in order of decreasing correlation coefficient with \NH. In Table 5 we
give the mutual correlation coefficients between all pairs of DIBs in this study.
Of the 28 pairs, 27 have correlations greater than 0.9. The exception is
$\lambda\lambda$5797.1--5487.7, with $r = 0.87.$
DIBs which arise from the same carriers, or whose carriers may have been formed
in the presence of a third, common carrier, would have correlation coefficients very close to unity.
None of the pairs considered here have such a high correlation, with the exception of
\lam\lam6196.0--6613.6, as discussed in Paper IV.

We now consider correlations between $\lambda5780.5$ and the other seven DIBs.
Correlation coefficients are given in Table 6, with the modified set, which excludes
the three outlier sight lines, in column 2, and the full set in column 5. Note that
correlations for the full and modified set are identical to within our errors. This is
not true for the full and modified set for the relation between the DIBs and \NH\
(Table 2). Table 6 also gives the coefficients of the best
fit lines for the DIB--$W_{\lambda}$(5780.5) plots, excluding the three outliers.

\begin{figure}
\epsscale{1.0}
\plotone{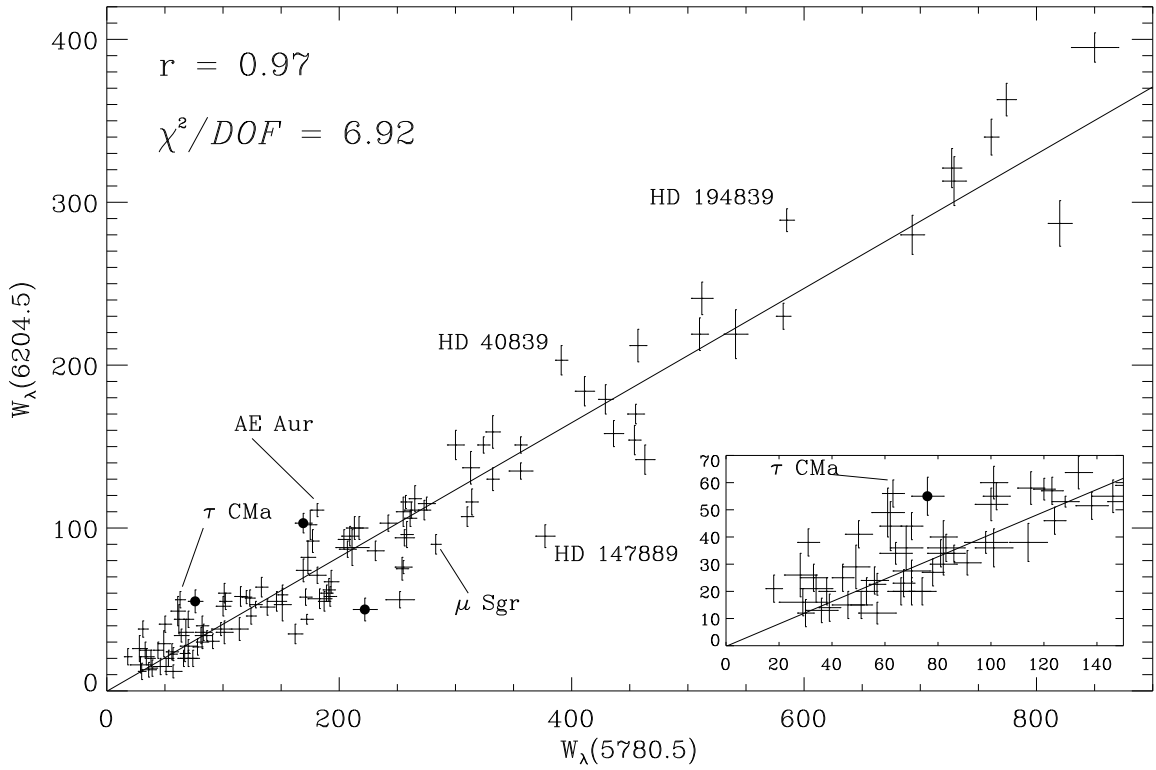}
\caption{$W_{\lambda}(6204.5)$ $vs.$ $W_{\lambda}(5780.5)$.
\label{6204_5780}}
\end{figure}

\begin{figure}
\epsscale{1.0}
\plotone{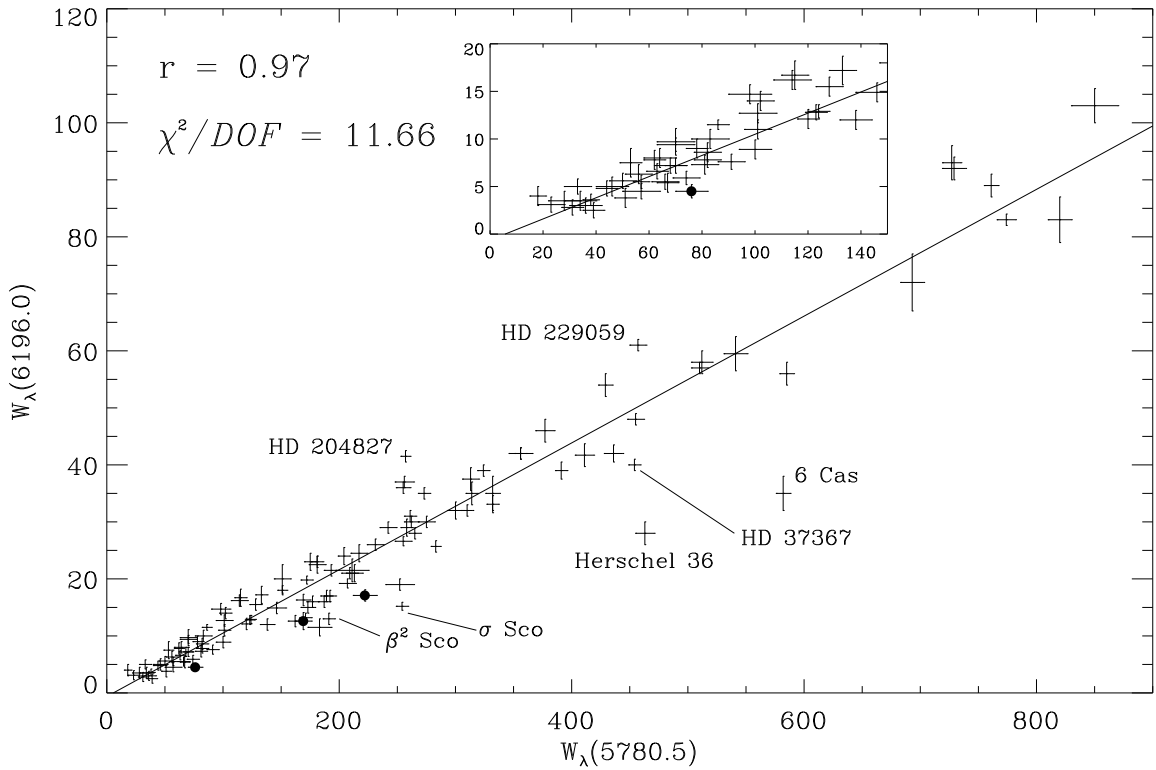}
\caption{$W_{\lambda}(6196.0)$ $vs.$ $W_{\lambda}(5780.5)$.
\label{6196_5780}}
\end{figure}

\begin{figure}
\epsscale{1.0}
\plotone{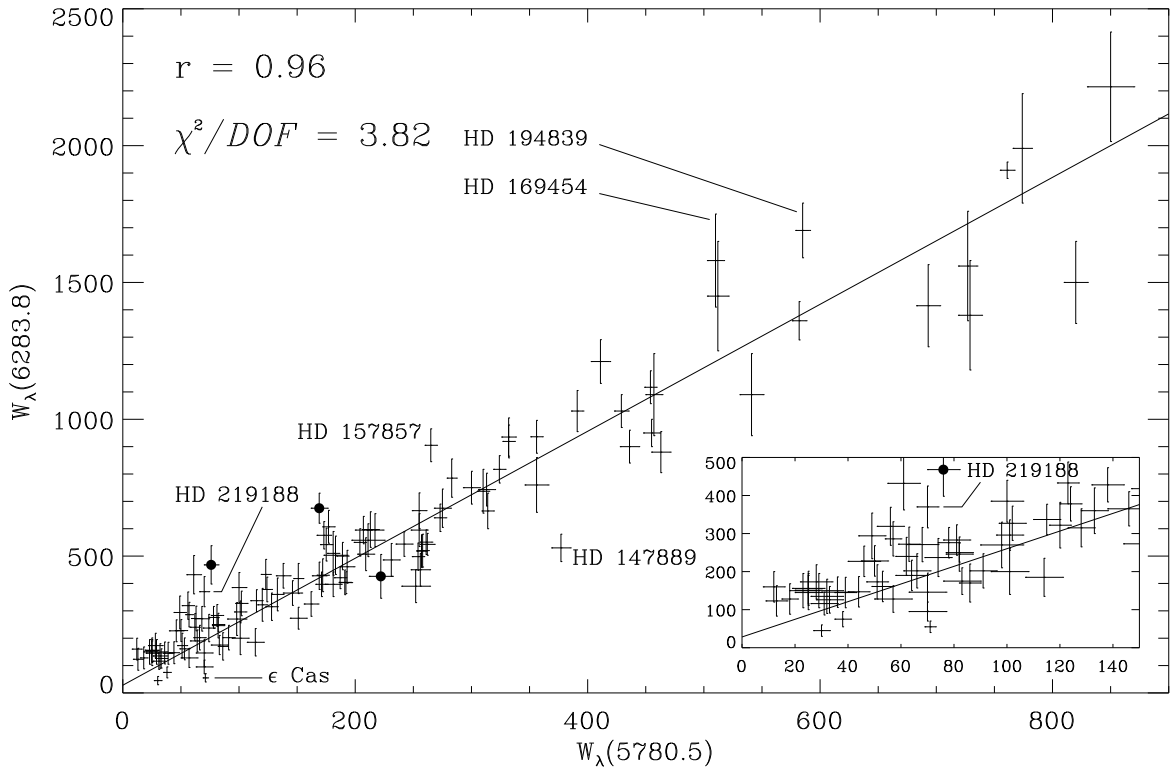}
\caption{$W_{\lambda}(6283.8)$ $vs.$ $W_{\lambda}(5780.5)$.
\label{6283_5780}}
\end{figure}

\begin{figure}
\epsscale{1.0}
\plotone{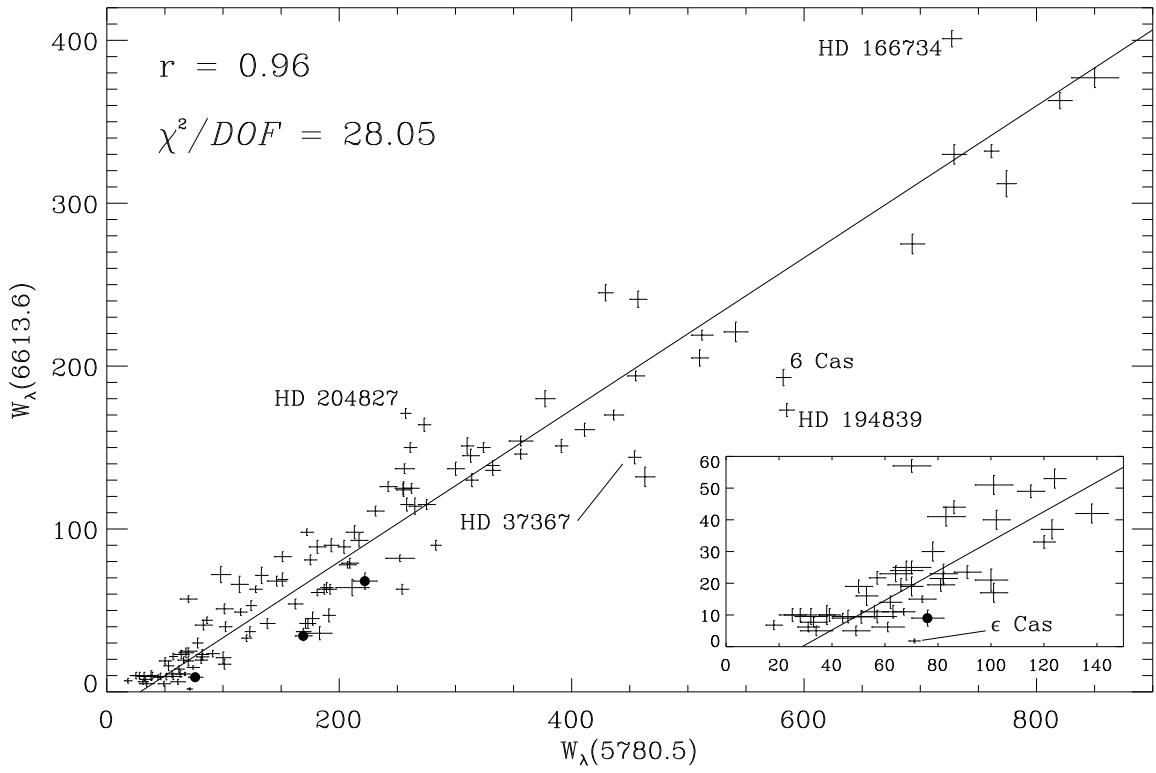}
\caption{$W_{\lambda}(6613.6)$ $vs.$ $W_{\lambda}(5780.5)$. This shows most
clearly the threshold effect -- some minimum amount of \lam5780.5 must be
present before \lam6613.8 begins to appear.
\label{6614_5780}}
\end{figure}

\begin{figure}
\epsscale{1.0}
\plotone{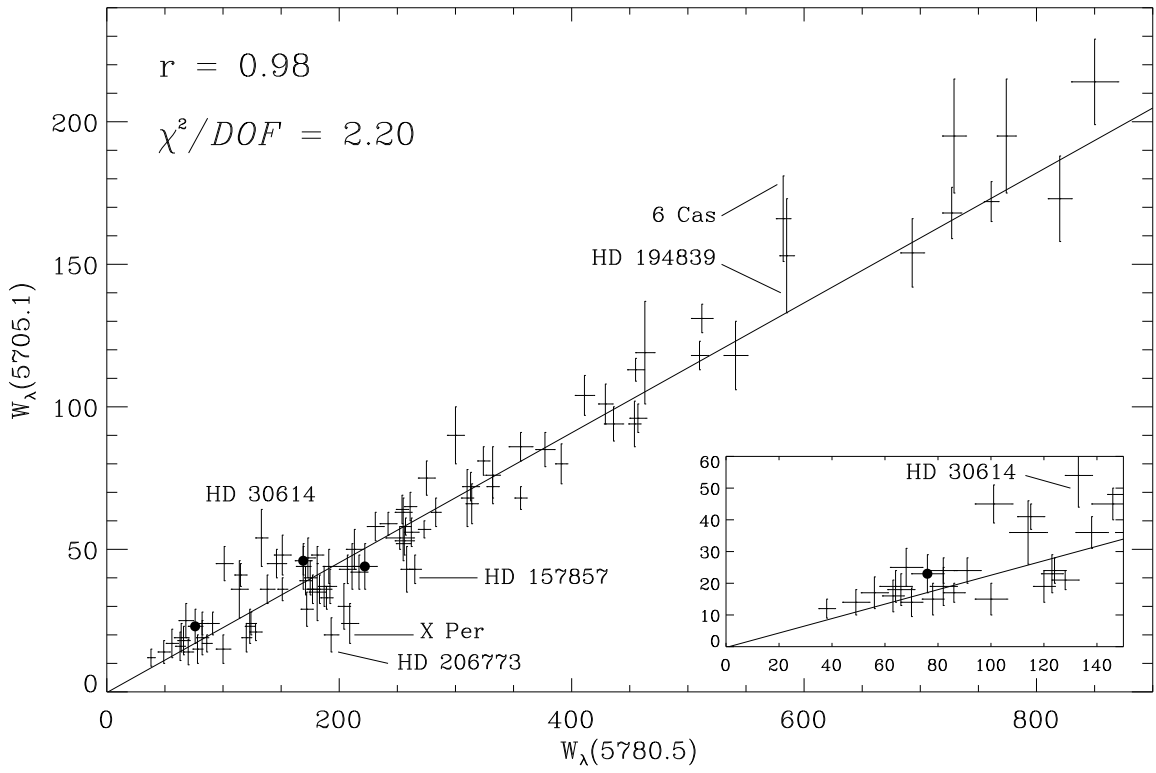}
\caption{$W_{\lambda}(5705.1)$ $vs.$ $W_{\lambda}(5780.5)$. This has the
smallest reduced $\chi^2$ and the highest correlation coefficient of the DIBs
with respect to \lam5780.5.
\label{5705_5780}}
\end{figure}

\begin{figure}
\epsscale{1.0}
\plotone{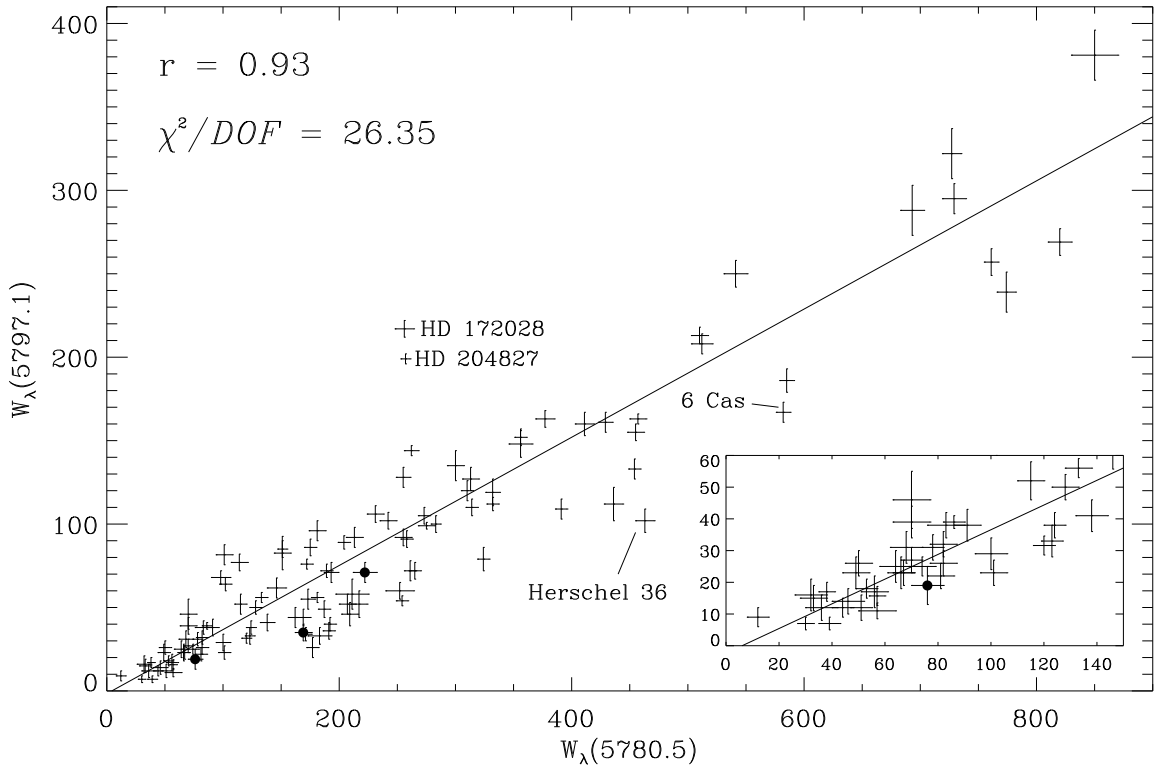}
\caption{$W_{\lambda}(5797.1)$ $vs.$ $W_{\lambda}(5780.5)$.
\label{5797_5780}}
\end{figure}

\begin{figure}
\plotone{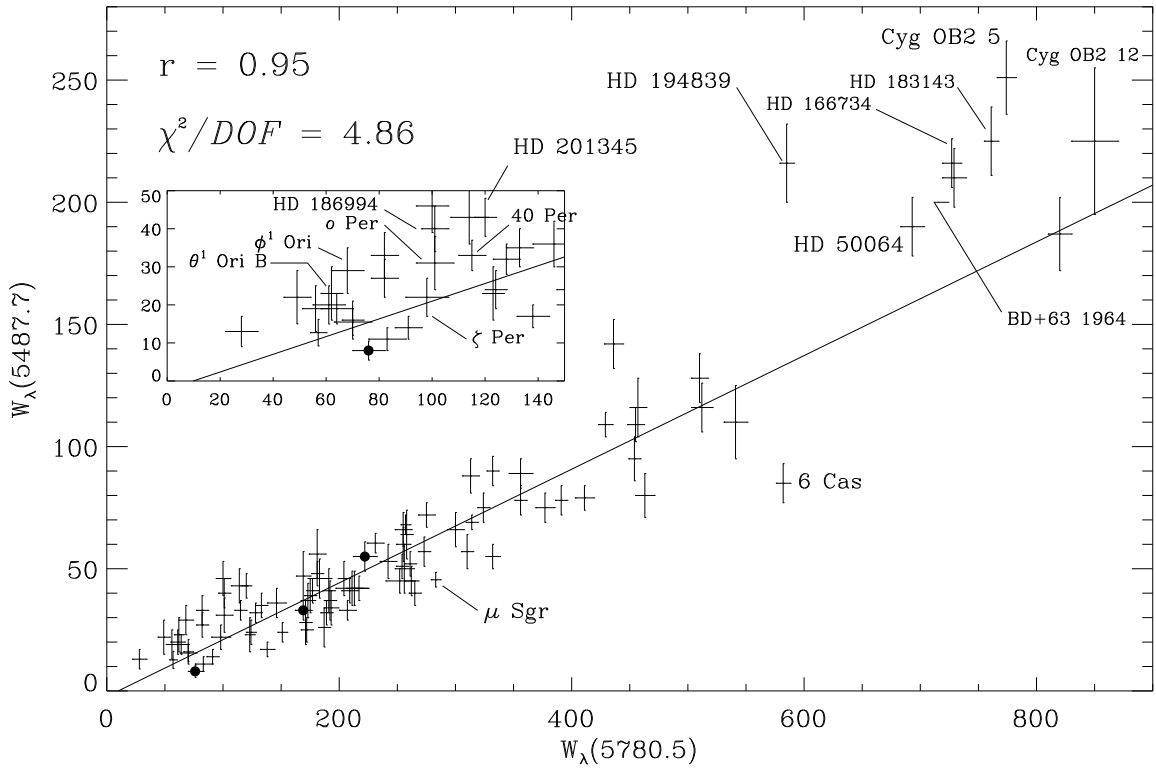}
\caption{$W_{\lambda}(5487.7)$ $vs.$ $W_{\lambda}(5780.5)$.
\label{5487_5780}}
\end{figure}

Figures~\ref{6204_5780} through \ref{5487_5780} show the relationship between
$\lambda5780.5$ and the other DIBs in our sample. We note here some of the
characteristics of these plots.

Figure~\ref{6204_5780} shows the $\lambda6204.5$ $vs.$ $\lambda5780.5$
relationship (correlation coefficient $r=0.97$). The greatest outliers in terms of
the number of standard deviations
off the best fit line in both the $x$ and $y$ directions combined, are HD 40839,
HD 194839, HD 147889, AE Aur, $\tau$ CMa, and $\mu$ Sgr. However, with
a correlation coefficient of 0.97, this is one of the best correlations observed, even
better than $\lambda$5780.5 with \NH\ (r=0.94).

Figure~\ref{6196_5780} shows the correlation of the DIBs $\lambda$6196.0 and
$\lambda$5780.5 ($r=0.97$).  The outliers here include 6 Cas, HD 204827,
$\sigma$ Sco, Herschel 36, HD 37367, HD 229059, and $\beta^2$ Sco.
The reduced $\chi^2$ is almost 12 (Table 6), indicating a rather large scatter
of data points, even though they are highly correlated.

The $\lambda6283.8$ $vs.$ $\lambda5780.5$ correlation ($r=0.96$) in Figure~\ref{6283_5780} has
only two points that are more than $5\sigma$ off the best fit line in both the $x$ and
$y$ directions, $\epsilon$ Cas and HD 147889. Other outliers include HD 157857, HD 169454,
HD 194839, and HD 219188.  \lam6283.8 is by far the strongest
of the eight DIBs but there is no evidence of saturation in  Figure~\ref{6283_5780}.

Figure~\ref{6614_5780} plots $\lambda$6613.6 $vs.$  $\lambda$5780.5 ($r=0.96$).  This relation
has a reduced $\chi^2$ of more than 28, the highest of all DIB-$W_{\lambda}$(5780.5)
pairs (Table 6). Among the points which deviate the most are  HD 204827,
HD 194839, $\epsilon$ Cas, HD 37367, 6 Cas, and HD 166734.

Figure~\ref{5705_5780} plots $\lambda5705.1$ $vs.$ $\lambda5780.5$ ($r=0.98$). This relation
has the highest correlation coefficient and the lowest scatter ($\chi^2 = 2.2$) of any of
our DIB pairs. No point deviates from the best fit line in the
$W_{\lambda}(5705.5)$ direction by more than $4\sigma.$ The largest outliers are
6 Cas, HD 30614, HD 206773, HD 157857, X Per, and HD 194839.
The correlation coefficient is comparable to that of the \lam\lam6196.0$-$6613.6 pair
discussed in Paper IV, and will be subject to additional, detailed study by our group.

The $\lambda$5797.1 $vs.$ $\lambda$5780.5 plot ($r=0.92$), Figure~\ref{5797_5780}, also
exhibits high scatter, with reduced $\chi^2$ exceeding 26. We
notice a group of points falling below the line near $W_{\lambda}(5780.5) = 175$ m\AA.
This includes HD 53975 at $(x,y)$ = (177,26), HD 37903 at (183,33), $\beta^2$ Sco
at (191,36), $\omega^1$ Sco at (192,40), $\nu$ Sco AB
at (187,49),  and $\beta^1$ Sco AB at (171,34). Other outliers include HD 204827,
HD 172028, Herschel 36, and 6 Cas.

The $\lambda$5487.7 vs.  $\lambda$5780.5 plot ($r=0.95$), shown in Figure~\ref{5487_5780},
exhibits a group of points sitting above the line at both the lowest and the highest
column densities. The first group includes HD 201345 at (100,46), HD 186994 at
(101,40), $\zeta$ Per at (98,22), $o$ Per at (101,31), 40 Per at (115,33),
$\phi^1$ Ori at (68,29), and $\theta^1$ Ori B at (61,20). The second group includes
HD 194839, HD 166734, HD 183143, Cyg OB2 5, Cyg OB2 12, BD+63 1964,
and HD 50064. The two largest outliers below the line are 6 Cas
and $\mu$ Sgr. This is the only DIB-DIB plot for which there
is a possible systematic deviation from a linear relationship at high column densities.
In this case \lam5780.5 may be saturating at $W_{\lambda}\ga600$m\AA. However, no saturation
is indicated at similar line strengths in the correlations with other DIBs, so it is unlikely
that this is the cause of this distribution of points. If this is correct, it implies that \lam5487.7
is getting stronger per H atom for high levels of \lam5780.5.

\section{Discussion}

Most interstellar quantities will show a positive correlation with each other simply
due to the increase of interstellar material with distance. This is most clearly reflected
in the correlation of DIBs with extinction. For the eight DIBs considered here the
correlation coefficients with \ebv\ range from $0.80-0.85$ (Table 4), with an average of 0.82.
Thus, we may regard r $\sim0.86-0.88$ as the minimum required to indicate that
two quantities are physically correlated at a significant level.

By this measure $\lambda5780.5$ is the only DIB in our study that is unambiguously
well-correlated with \NH, and even in this case the outlier points
(Figure~\ref{nh_5780}) are prominent exceptions. These three stars have in
common the presence of strong local radiation fields. Noticing that the outliers have
particularly low abundances of $\lambda$5780.5 relative to their H column
densities compared to the general correlation, a possible explanation is that
the radiation fields in these regions are hard enough to destroy the carrier of
$\lambda$5780.5, and that local radiation generally regulates DIB
abundances, which are likely characterized by distinct critical
wavelengths of ionizing radiation. Because all of the DIBs are well correlated with
$\lambda$5780.5, with correlation coefficients $0.92-0.97$ (Table 5),
but only $\lambda$5780.5 is very well correlated with \NH\ (Table 2),
we infer that the critical energy for regulating $\lambda$5780.5 is
similar to the energy that regulates the H abundance, and greater than
that of the other DIBs. We note that the destruction mechanism may be
simple ionization, but it could also be dissociation. However, the
apparent destruction effect in the presence of significant photon
fluxes indicates that the $\lambda$5780.5 carrier might be an ion.

These conclusions are supported by Sonnentrucker at al. (1997), who
examined the ionization properties of the DIBs $\lambda\lambda$5780.5, 5797.1,
6379.3, and 6613.6. They conclude that the carriers of these DIBs are 
separate gas-phase molecules. By comparing W$_\lambda$/\ebv\ as a function
of \ebv\ for each DIB, they find that $\lambda5780.5$ reaches its maximum at
the lowest value of \ebv, indicating that the carrier of this DIB is the most resistant
of the four DIB carriers to strong UV fields.

While the main interstellar clouds in the three outlier sight lines are subject to higher
than average radiation fields, they are not the only stars in our sample
with this property. The \NaI\ and \KI\ lines in the two Trapezium sight lines,
$\theta^1$ Ori C and HD 37061, are exceptionally weak, relative to \NH. $\rho$ Oph A,
however, is a bit puzzling. Various Sco-Oph sight lines (and some others
in other regions) exhibit relatively weak \NaI\ and \KI\ (Welty \& Hobbs 2001).
For these targets only $\rho$ Oph A is among the most discrepant in \lam5780.5
$vs.$ \NH, but most other Sco-Oph sight lines tend to be deficient in \lam5780.5, as well.
Unfortunately, we do not have a good quantitative measure of
radiation field strength in many cases. One can use the higher \htwo\ rotational
level populations (J=4,5) to estimate this, but the column densities of those levels are
often very hard to determine accurately because the lines are generally on the flat part of
curve of growth for most sight lines of interest here. We have no measure of \htwo\
toward HD 37061 and the higher J lines
have not been reported toward $\rho$ Oph A. The error on \Nhtwo\ toward
$\theta^1$ Ori C is among the largest in our sample. Obtaining more extensive data
on the strength of the radiation fields along a large number of sight lines
would be a very interesting study.

If two DIBs are formed from transitions between a single
ground state and two different vibronic levels, their measured strengths should
be perfectly correlated, with a correlation coefficient of unity. This is not quite
true even for the best observed DIB$-$DIB correlation, $\lambda6196.0$ $vs.$
$\lambda6613.6.$ McCall et al. (Paper IV) explore this particular
pair and discuss the possibility that the true errors are underestimated.
The most likely causes of such a situation are: 1) there are errors
in continuum placement; 2) there are blends of DIBs that are not physically related;
and 3) there is uncorrected contamination from telluric water vapor lines.
The first of these could be especially true for $\lambda5780.5$, which is
superimposed on a much wider feature centered at 5778 \AA\  (Herbig 1975; see also
the combined spectral plots shown in Figure 11 of Papers II and III).
Without knowing the origin of the features that make up this blend
(Krelowski, Schmidt, \& Snow 1997), it is hard to
evaluate this effect. For the much narrower
$\lambda5797.1$ DIB the second effect may be operative. The feature is multiple
and the line shapes differ from star to star in our data (see Figure~\ref{dib8panel}).
We suspect the short wavelength contribution to this DIB is a \ctwo\
DIB (Paper I), while the long
wavelength component is more closely related to \lam5780.5. This explanation
is consistent with the well-known $\sigma-\zeta$ effect of 5780.5 and 5797.1:
in the prototype ``$\sigma$'' sight line, $\sigma$ Sco, \lam5780.5 is much
deeper than \lam 5797.1, while in the prototype ``$\zeta$'' sight line,
$\zeta$ Oph, the depths of the two lines are comparable (Krelowski \& Sneden
1995). The stars with high \Nhtwo\ and low \Wlam(5780) are all $\zeta-$type
stars, including strong \ctwo\ DIBs, and produce a deep blueward component
of the proposed \ctwo\ DIB at 5797.1\AA.

The $\sigma-\zeta$ effect might be more of a geometrical effect than an
ionization effect. It is well-known that \lam5780.5 is poorly correlated with trace
neutral species, such as \KI\ and \NaI\ (Welty et al. 2006). This is also true of other
$\sigma-$ type DIBs that are best correlated with \NH, such as \lam\lam 6283.8,
6204.5. On the other hand, the $\zeta-$type DIB \lam5797.1 is better correlated with the
trace neutrals (Galazutdinov et al. 2004). These neutrals predominantly exist in the cores
of clouds, which provide adequate shielding of the UV field, and where high molecular
fractions of \htwo\ exist due to self-shielding. These results are consistent with
the $\sigma-$type DIBs existing in the outer regions of clouds and the $\zeta-$type
DIBs existing deep in the interior of the clouds. The variable strength ratio of
\lam\lam5780.5, 5797.1 has been discussed by several authors, such as Krelowski,
Schmidt, \& Snow (1997). \lam5797.1 has the lowest correlation coefficient ($r=0.93$)
with \lam5780.5 of any of the DIBs in our study (Table 6), even though it still exceeds
the value expected from the general growth of interstellar material, as noted above.

Most of our sight lines penetrate multiple clouds and therefore average out these effects.
Cami et al. (1997) studied 13 stars with only single clouds along the lines of sight. They
confirm the classification of DIBs into four different families, including $\sigma$ and $\zeta.$
However, the most convincing evidence of this hypothesis will come from mapping
several clouds spatially by observing multiple stars at various locations behind the clouds. Correlations
among DIBs and with neutral atoms and with molecules, such as \htwo, \ctwo, CN, and CH will
indicate whether the geometrical interpretation is correct. This is the subject of a future
study by our group.

Examination of DIBs that fall in the clean parts of the spectra of a number of stars
indicate the presence of a number of interstellar absorption components.
Galazutdinov \etal\ (2002a, 2005) argue that
most strong DIBs have structure. While these authors
conclude the structures are evidence of the R, P, Q rotational structure of
molecules of modest size, our study suggests that blends of unrelated
DIBs may be the origin of structure in some cases.

The question of what features in the neighborhood of the main DIB to include in
the measurements of equivalent width is difficult. We do not have the fine spectral
resolution of Galazutdinov et al. (2002a) and cannot separate the blended features
in two of our DIBs, \lam5797.1 and \lam6204.5, but we have a larger dataset at
very high S/N. It is difficult even at high resolution to conclude whether lines have
structure due to multiple components along a line of sight, blends of unrelated
species, or structure from energy levels in a common carrier. Even if
we know in advance that DIBs from two different carriers are involved, we still do not
know what profiles to use in the deconvolution fitting process. We therefore measure the
full EW of the blends. Figure~\ref{dib8panel} shows that the two features in the
profile of \lam6204.5 approximately scale with each other, and this DIB has among
the highest correlation coefficients with the other DIBs (Table 5) and with \NH\ (Table 2),
and the correlation with \lam5780.5 has low $\chi^2$ (Table 6). In contrast,
the features of \lam5797.1 do not scale with each other, this DIB has low correlation
coefficients with the other DIBs and \NH, and the correlation with \lam5780.5 has
very high $\chi^2$. It would appear that the two features blended in the \lam6204.5
feature we measure as one feature are either from the same carrier or from two
separate carriers that are almost perfectly correlated. As previously noted, such
considerations may help guide the correct placement of limits in future studies.

Examination of Table 6 shows that the intercepts of the best fit lines pass
through the origin for the correlation of $\lambda5780.5$ with DIBs
$\lambda\lambda$5705.1, 6204.5, 5487.7, and possibly $\lambda\lambda$5797.1
and 6196.0. However, our data show that this is not true for \lam6283.8, and
especially \lam6613.6. This may be evidence of a threshold effect,
such that a substantial amount of 5780.5 must be produced before 6613.6 can begin
to form. Since \NH\ is very well correlated with $\lambda5780.5$, this implies that
some minimum column density of H is required before the appearance of some
DIBs becomes evident.

Figure~\ref{6283_5780} shows a possible threshold in the opposite
sense. This is the only case in which the linear fit has a statistically significant
positive intercept, indicating that 6283.8 appears before 5780.5. In addition,
below $W_{\lambda}(5780.5) \approx 150$\,m\AA\ there are significantly more
points above the best fit line than below. Below  $W_{\lambda}(5780.5)
\approx 50 $\,m\AA\ the distribution flattens, with $W_{\lambda}(6283.8)$
remaining approximately constant as $W_{\lambda}(5780.5)$ decreases.
This may be caused by two effects: 1) 6283.8 is among the broadest DIBs in this study
(FWHM = 4.77\,\AA, Table 2),
making continuum fitting somewhat less reliable than for narrower DIBs.
2) There is considerable telluric contamination in such a wide DIB and our
nightly blanket correction for telluric lines (\S2) may introduce unrecognized errors. Indeed,
we estimate our minimum detectable DIB equivalent width to be approximately
150\,m\AA, about the same as the plateau in Fig.~\ref{6283_5780}. This line
is asymmetric and far from Gaussian in shape, so it would not be evident from
an inspection of the line profile if one or more interfering lines are present.

One of the most useful results of this work is the ability to estimate the total column
density of atomic hydrogen along a Galactic sight line based on a measurement of the equivalent
width of a single DIB. The tight correlation between \NH\ and \Wlam(5780.5) shown in
Figure~\ref{nh_5780} demonstrates that this technique can be used in most
cases, and in fact this has a higher correlation coefficient and lower $\chi^2$ than
\ebv $\,vs.$ \NH. However, the relationship fails to hold for the outlier stars, so care
must be taken if the sight line passes through a region of high UV-radiation, assuming
this influences the abundance of the DIB carriers, as discussed above. A second
empirical application is to compare the correlations presented here with correlations
among the same features in other galaxies, in hopes that differences can be related
to different physical properties of the galaxies and lead to an explantion of the DIBs.
For example, note that since \NCH\ and \Nhtwo\ are moderately well correlated
(Welty et al. 2006), the total H (atomic plus molecular) column density can be
estimated for stars too faint to have measured far UV extinction curves.

Welty et al. (2006) examined the strengths of the $\lambda\lambda$5780.5, 5797.1, and
6283.8 DIBs toward a relatively small number of stars in the Magellanic Clouds.
They found that the correlations of these with \NH\
were lower than with \ebv, which is not true for most of the DIBs in the current study.
They also found that these DIBs are systematically weaker relative to \NH\ than they are in
the Milky Way by factors of $7-9$ (in the LMC) and $\sim20$ (in the SMC),
and weaker by about a factor of 2 relative to \ebv. In a slightly larger sample of stars
with improved \NH\ for some sight lines,
Welty et al. (in preparation) find that the correlations are slightly better than
those reported in the earlier study but still not as good as ours for Galactic sight lines.
These differences may be due to lower metallicities or stronger radiation fields found
in the clouds; see Welty et al. (2006) for additional discussion of these points.

\section{Summary}

We have used a large database of high signal to noise ratio spectra of 133 stars to 
perform one of the most extensive comparisons to date between strengths of DIB
pairs and between DIBs and \NH, \Nhtwo, and \ebv. We have presented linear fit
parameters and correlation coefficients for these relationships. We reach the following
conclusions.

\begin{enumerate}

\item Only one DIB in our study, $\lambda$5780.5, is unambiguously well-correlated with \NH, in
the sense that the correlation coefficient exceeds what one would expect from the
growth of interstellar material with distance.

\item None of the DIB--DIB correlation coefficients considered here are high enough
to conclude that any pair arises from the same carrier. However, as described in
Paper IV, the \lam\lam6196.0$-$6613.6 pair may be perfectly correlated if the errors
were underestimated by a small amount. The correlation of the \lam\lam5780.5$-$5705.1
DIBs is also very high and further study is warranted.

\item Seven of the eight DIBs, excepting \lam5487.7, are better correlated with
\NH\ than with \ebv.

\item All eight DIBs are very poorly correlated with \Nhtwo. Even when we restrict
the sight lines to those with log[\Nhtwo] $>$ 18, the correlations are poor. At a single
value of $W_{\lambda}$(5780.5) the column density of \htwo\ can vary by a factor
of $10^5$. This occurs just at the level of reddening corresponding to the formation
of enough \htwo\ to permit self-shielding.

\item The excellent correlation of \lam5780.5 $vs.$ \NH\ may be understood if the
critical energy of radiation needed to ionize the two species is similar. The greater
the flux of H ionizing radiation, the higher the degree of ionization of the \lam5780.5
DIB carrier.

\item Most of the linear fits to the DIB--DIB correlations pass through the origin. This is not
true for \lam\lam 6283.8 and 6613.6. This may be due to continuum placement errors,
the presence of interfering DIBs, improperly corrected telluric contamination, or
a threshold effect, in which one DIB cannot form until a significant amount of another
DIB is present.

\item One of the most practical uses of the results presented here is the ability to
estimate \NH\ in Galactic sight lines based on a measurement of the equivalent width
of \lam(5780.5). One must be careful to exclude sight lines in high radiation
environments since these maybe responsible for the outliers in this otherwise tight
relationship. In addition, the correlations presented here may be compared to correlations
found in other galaxies, and this may help identify the carriers of DIBs.

\end{enumerate}

\acknowledgments

This work is based on observations obtained with the Apache Point 3.5\,m telescope,
which is owned and operated by the Astrophysical Research Consortium.
We thank Tom Fishman for help with an early version of this paper and T. Oka for many
useful conversations and insights into the nature of DIBs. T.P.S. was supported by NASA
grant NNX08AC14G. B.J.M. gratefully acknowledges support from the David and Lucile
Packard Foundation and the University of Illinois.


\tabletypesize{\tiny}
\begin{landscape}
\begin{deluxetable}{lccccccccccccc}
\tablenum{1}
\tablecolumns{1}
\tablewidth{0pt}
\tablecaption{Stellar Properties, Interstellar Line Data, and DIB Equivalent Widths (m\AA)}
\tablehead{
\colhead{HD} &
\colhead{Sp Type} &
\colhead{\ebv\tablenotemark{a}} &
\colhead{log(\NH)} &
\colhead{log(\Nhtwo)} &
\colhead{References\tablenotemark{b}} &
\colhead{5487.7} &
\colhead{5705.1} &
\colhead{5780.5} &
\colhead{5797.1} &
\colhead{6196.0} &
\colhead{6204.5} &
\colhead{6283.8} &
\colhead{6613.6}
}
\startdata

2905	 & 	B1Iae	 & 	0.33	 & 	21.26	 $\pm$ 	0.09	 & 	20.27	 $\pm$ 	0.09	 & 	3,6	&	69 $\pm$ 3	 & 	66 $\pm$ 7	 & 	314 $\pm$ 5	 & 	110 $\pm$ 5	 & 	35 $\pm$ 2	 & 	116 $\pm$ 8	 & 	665 $\pm$ 65	 & 	130 $\pm$ 4	 \\
10516	 & 	B2Vep	 & 	0.20	 & 				 & 	19.08	 $\pm$ 	0.09	 & 	--,6	&	$<$  10	 & 	$<$  15	 & 	67 $\pm$ 4	 & 	23 $\pm$ 4	 & 	5.4 $\pm$ 1	 & 	23 $\pm$ 5	 & 	$<$  200	 & 	11 $\pm$ 1	 \\
11415	 & 	B3III	 & 	0.05	 & 				 & 				 & 		&	$<$  20	 & 	$<$  6	 & 	71 $\pm$ 2	 & 	$<$  8	 & 	$<$  2	 & 	$<$  12	 & 	55 $\pm$ 15	 & 	1.8 $\pm$ 0.6	 \\
16219	 & 	B5V	 & 	0.04	 & 				 & 				 & 		&	$<$  15	 & 	$<$  12	 & 	36 $\pm$ 6	 & 	12 $\pm$ 4	 & 	3 $\pm$ 0.8	 & 	13 $\pm$ 4.5	 & 	146 $\pm$ 40	 & 	$<$  6	 \\
19374	 & 	B1.5V	 & 	0.13	 & 	21.06	 $\pm$ 	0.11	 & 				 & 	1,--	&	17 $\pm$ 3	 & 	36 $\pm$ 5	 & 	138 $\pm$ 6	 & 	41 $\pm$ 5	 & 	12 $\pm$ 1	 & 	51.5 $\pm$ 5	 & 	428 $\pm$ 45	 & 	42 $\pm$ 3	 \\
20041	 & 	A0Ia	 & 	0.72	 & 				 & 				 & 		&	109 $\pm$ 5	 & 	101 $\pm$ 7	 & 	429 $\pm$ 6	 & 	161 $\pm$ 6	 & 	54 $\pm$ 2	 & 	179 $\pm$ 9	 & 	1030 $\pm$ 60	 & 	245 $\pm$ 5	 \\
21071	 & 	B7V	 & 	0.05	 & 				 & 				 & 		&	14 $\pm$ 3	 & 	24 $\pm$ 4	 & 	91 $\pm$ 5	 & 	38 $\pm$ 5	 & 	7.6 $\pm$ 0.8	 & 	30.5 $\pm$ 4.5	 & 	202 $\pm$ 45	 & 	23.5 $\pm$ 2	 \\
21483	 & 	B3III	 & 	0.56	 & 				 & 				 & 		&	56 $\pm$ 10	 & 	35 $\pm$ 10	 & 	181 $\pm$ 7	 & 	96 $\pm$ 6	 & 	22.5 $\pm$ 1.5	 & 	71 $\pm$ 5	 & 	397 $\pm$ 45	 & 	89 $\pm$ 4	 \\
21389	 & 	A0Iae	 & 	0.57	 & 				 & 				 & 		&	79 $\pm$ 5	 & 	104 $\pm$ 7	 & 	411 $\pm$ 8	 & 	160 $\pm$ 7	 & 	41.7 $\pm$ 2	 & 	184 $\pm$ 9	 & 	1211 $\pm$ 80	 & 	161 $\pm$ 4	 \\
22951 (40 Per)	 & 	B0.5V	 & 	0.27	 & 	21.04	 $\pm$ 	0.11	 & 	20.46	 $\pm$ 	0.09	 & 	5,6	&	33 $\pm$ 4	 & 	41 $\pm$ 4	 & 	115 $\pm$ 5	 & 	52 $\pm$ 6	 & 	16.7 $\pm$ 1.5	 & 	58 $\pm$ 6	 & 	337 $\pm$ 40	 & 	49 $\pm$ 2	 \\
23180 ($o$ Per)	 & 	B1III	 & 	0.31	 & 	20.82	 $\pm$ 	0.09	 & 	20.61	 $\pm$ 	0.09	 & 	1,6	&	31 $\pm$ 7	 & 	45 $\pm$ 6	 & 	101 $\pm$ 7	 & 	81.6 $\pm$ 6	 & 	12.7 $\pm$ 1	 & 	36 $\pm$ 7	 & 	200 $\pm$ 60	 & 	51 $\pm$ 3	 \\
281159	 & 	B5V	 & 	0.85	 & 	21.38	 $\pm$ 	0.3	 & 	21.09	 $\pm$ 	0.19	 & 	9,9	&	57 $\pm$ 7	 & 	68 $\pm$ 10	 & 	310 $\pm$ 5	 & 	120 $\pm$ 6	 & 	32 $\pm$ 1	 & 	107 $\pm$ 6	 & 	737 $\pm$ 80	 & 	151 $\pm$ 5	 \\
23408	 & 	B8III	 & 	0.02	 & 				 & 	19.75	 $\pm$ 	0.13	 & 	--,6	&	$<$  10	 & 	$<$  15	 & 	$<$  25	 & 	$<$  5	 & 	2.7 $\pm$ 0.8	 & 	$<$  9	 & 	$<$  120	 & 	7 $\pm$ 1.5	 \\
23480	 & 	B6IVe	 & 	0.08	 & 				 & 	20.11	 $\pm$ 	0.09	 & 	--,6	&	$<$  10	 & 	$<$  15	 & 	31 $\pm$ 5	 & 	$<$  5	 & 	$<$  2.5	 & 	$<$  10	 & 	116 $\pm$ 30	 & 	$<$  4	 \\
24398 ($\zeta$ Per)	 & 	B1Ib	 & 	0.31	 & 	20.8	 $\pm$ 	0.08	 & 	20.68	 $\pm$ 	0.09	 & 	1,6	&	43 $\pm$ 7	 & 	36 $\pm$ 10	 & 	114 $\pm$ 7	 & 	77 $\pm$ 5	 & 	16.2 $\pm$ 1	 & 	38 $\pm$ 7	 & 	185 $\pm$ 50	 & 	66 $\pm$ 5	 \\
24534 ( X Per)	 & 	O9.5pe	 & 	0.59	 & 	20.73	 $\pm$ 	0.06	 & 	20.92	 $\pm$ 	0.04	 & 	1,8	&	22 $\pm$ 5	 & 	$<$  30	 & 	98 $\pm$ 8	 & 	68 $\pm$ 4	 & 	14.7 $\pm$ 1	 & 	38 $\pm$ 4	 & 	270 $\pm$ 60	 & 	72 $\pm$ 5	 \\
24760	 & 	B0.5V+A2	 & 	0.10	 & 	20.45	 $\pm$ 	0.11	 & 	19.52	 $\pm$ 	0.13	 & 	1,6	&	$<$  12	 & 	$<$  20	 & 	81 $\pm$ 5	 & 	22 $\pm$ 4	 & 	7.3 $\pm$ 1	 & 	34 $\pm$ 5	 & 	283 $\pm$ 40	 & 	19.5 $\pm$ 2	 \\
24912	 & 	O7e	 & 	0.33	 & 	21.05	 $\pm$ 	0.08	 & 	20.54	 $\pm$ 	0.08	 & 	1,6	&	41 $\pm$ 5	 & 	24 $\pm$ 7	 & 	209 $\pm$ 7	 & 	46 $\pm$ 7	 & 	21 $\pm$ 1	 & 	95 $\pm$ 6	 & 	507 $\pm$ 60	 & 	79 $\pm$ 3	 \\
26571	 & 	B9IIIp	 & 	0.25	 & 				 & 				 & 		&	$<$  15	 & 	48 $\pm$ 7	 & 	151 $\pm$ 7	 & 	82.6 $\pm$ 10	 & 	20 $\pm$ 2.5	 & 	53 $\pm$ 10	 & 	273 $\pm$ 40	 & 	83 $\pm$ 3	 \\
27778	 & 	B3V	 & 	0.37	 & 				 & 	20.79	 $\pm$ 	0.06	 & 	8,8	&	$<$  20	 & 	17 $\pm$ 3	 & 	86 $\pm$ 4	 & 	39 $\pm$ 2	 & 	11.5 $\pm$ 0.5	 & 	34 $\pm$ 3	 & 	170 $\pm$ 50	 & 	44 $\pm$ 2	 \\
28375	 & 	B3V	 & 	0.10	 & 				 & 				 & 		&	$<$  12	 & 	18 $\pm$ 5	 & 	66 $\pm$ 5	 & 	23 $\pm$ 5	 & 	5.5 $\pm$ 0.8	 & 	20 $\pm$ 5	 & 	202 $\pm$ 45	 & 	19.5 $\pm$ 2	 \\
28497	 & 	B1Ve	 & 	0.03	 & 	20.23	 $\pm$ 	0.1	 & 	15.09	 $\pm$ 	0.1	 & 	3,11	&		 & 		 & 	13 $\pm$ 4	 & 	$<$  9	 & 	$<$  3	 & 	$<$  15	 & 	$<$  120	 & 	$<$  5	 \\
29647	 & 	B8IIIp	 & 	1.00	 & 				 & 				 & 		&	$<$  12	 & 	$<$  24	 & 	70 $\pm$ 7	 & 	39 $\pm$ 5	 & 	9.7 $\pm$ 1.4	 & 	20 $\pm$ 5	 & 	95 $\pm$ 25	 & 	57 $\pm$ 2	 \\
30614	 & 	O9.5Iae	 & 	0.30	 & 	20.97	 $\pm$ 	0.09	 & 	20.34	 $\pm$ 	0.08	 & 	1,6	&	35 $\pm$ 5	 & 	54 $\pm$ 10	 & 	133 $\pm$ 5	 & 	56 $\pm$ 3	 & 	17.2 $\pm$ 1.5	 & 	63.7 $\pm$ 6	 & 	360 $\pm$ 60	 & 	71.5 $\pm$ 5	 \\
34078 (AE Aur)	 & 	O9.5Ve	 & 	0.52	 & 	21.2	 $\pm$ 	0.11	 & 				 & 	1,--	&	48 $\pm$ 5	 & 	48 $\pm$ 3	 & 	181 $\pm$ 5	 & 	56 $\pm$ 3	 & 	23 $\pm$ 1	 & 	111 $\pm$ 4	 & 	510 $\pm$ 80	 & 	61 $\pm$ 2	 \\
34503	 & 	B5III	 & 	0.05	 & 				 & 				 & 		&		 & 		 & 	25 $\pm$ 5	 & 	$<$  12	 & 	$<$  3	 & 	$<$  15	 & 	145 $\pm$ 45	 & 	$<$  6	 \\
34798	 & 	B3V	 & 	0.04	 & 				 & 				 & 		&		 & 		 & 	12 $\pm$ 4	 & 	9 $\pm$ 3	 & 	$<$  5	 & 	$<$  25	 & 	160 $\pm$ 40	 & 	$<$  7	 \\
35149	 & 	B1V	 & 	0.11	 & 	20.56	 $\pm$ 	0.07	 & 	18.3	 $\pm$ 	0.11	 & 	1,9	&	12.7 $\pm$ 3.5	 & 	$<$  12	 & 	57 $\pm$ 3	 & 	15.7 $\pm$ 3	 & 	5.5 $\pm$ 0.7	 & 	22.6 $\pm$ 4	 & 	286 $\pm$ 45	 & 	21.7 $\pm$ 2	 \\
36512	 & 	B0V	 & 	0.04	 & 	20.27	 $\pm$ 	0.11	 & 				 & 	1,--	&		 & 		 & 	25 $\pm$ 5	 & 	$<$  9	 & 	$<$  12	 & 	$<$  25	 & 	155 $\pm$ 40	 & 	10 $\pm$ 2	 \\
36486	 & 	B0III+O9V	 & 	0.08	 & 	20.19	 $\pm$ 	0.05	 & 	14.74	 $\pm$ 	0.05	 & 	12,13	&		 & 		 & 	23 $\pm$ 5	 & 	$<$  15	 & 	3.1 $\pm$ 0.8	 & 	$<$  20	 & 	150 $\pm$ 45	 & 	$<$  5	 \\
36591	 & 	B1IV	 & 	0.07	 & 				 & 				 & 		&	13 $\pm$ 4	 & 	$<$  15	 & 	28 $\pm$ 6	 & 	$<$  10	 & 	3.5 $\pm$ 1	 & 	26 $\pm$ 8	 & 	173 $\pm$ 45	 & 	10 $\pm$ 2	 \\
36371	 & 	B5Iab	 & 	0.43	 & 				 & 				 & 		&	88 $\pm$ 7	 & 	72 $\pm$ 5	 & 	313 $\pm$ 7	 & 	127 $\pm$ 7	 & 	37.5 $\pm$ 2	 & 	137 $\pm$ 10	 & 	743 $\pm$ 60	 & 	145 $\pm$ 4	 \\
36822 ($\phi^1$ Ori)	 & 	B0III	 & 	0.14	 & 	20.84	 $\pm$ 	0.07	 & 	19.32	 $\pm$ 	0.07	 & 	1,6	&	29 $\pm$ 6	 & 	25 $\pm$ 6	 & 	68 $\pm$ 6	 & 	31 $\pm$ 5	 & 	7.2 $\pm$ 0.8	 & 	36 $\pm$ 8	 & 	271 $\pm$ 45	 & 	24 $\pm$ 3	 \\
36861	 & 	O8e	 & 	0.15	 & 	20.81	 $\pm$ 	0.12	 & 	19.12	 $\pm$ 	0.1	 & 	1,6	&		 & 		 & 	50 $\pm$ 5	 & 	26 $\pm$ 4	 & 	5.6 $\pm$ 0.8	 & 	41 $\pm$ 5	 & 	228 $\pm$ 40	 & 	19 $\pm$ 2	 \\
37021 ($\theta^1$ Ori B)	 & 	B0V	 & 	0.54	 & 				 & 				 & 		&	20 $\pm$ 5	 & 	$<$  18	 & 	61 $\pm$ 6	 & 	$<$  15	 & 	$<$  4	 & 	49 $\pm$ 9	 & 	432 $\pm$ 70	 & 	6.2 $\pm$ 1.5	 \\
37022 ($\theta^1$ Ori C)	 & 	O6	 & 	0.34	 & 	21.54	 $\pm$ 	0.11	 & 	15.65	 $\pm$ 	0.13	 & 	1,7	&	8 $\pm$ 2.5	 & 	23 $\pm$ 6	 & 	76 $\pm$ 6	 & 	19 $\pm$ 6	 & 	4.5 $\pm$ 0.7	 & 	55 $\pm$ 7	 & 	468 $\pm$ 70	 & 	9 $\pm$ 2.5	 \\
37043	 & 	O9III	 & 	0.07	 & 	20.2	 $\pm$ 	0.1	 & 	14.69	 $\pm$ 	0.11	 & 	1,7	&	$<$  15	 & 	$<$  30	 & 	34 $\pm$ 6	 & 	$<$  12	 & 	3.5 $\pm$ 1	 & 	21 $\pm$ 5	 & 	$<$  120	 & 	5 $\pm$ 1.5	 \\
37061	 & 	B1V	 & 	0.52	 & 	21.81	 $\pm$ 	0.07	 & 				 & 	3,--	&	33 $\pm$ 4	 & 	46 $\pm$ 5	 & 	169 $\pm$ 7	 & 	35 $\pm$ 5	 & 	12.6 $\pm$ 1.5	 & 	103 $\pm$ 6	 & 	675 $\pm$ 55	 & 	34.4 $\pm$ 3	 \\
37128	 & 	B0Iae	 & 	0.05	 & 	20.48	 $\pm$ 	0.11	 & 	16.28	 $\pm$ 	0.2	 & 	1,9	&		 & 		 & 	18 $\pm$ 3	 & 	$<$  5	 & 	4 $\pm$ 1	 & 	21 $\pm$ 5	 & 	128 $\pm$ 40	 & 	6.8 $\pm$ 1.4	 \\
37367	 & 	B2IV-V	 & 	0.40	 & 				 & 				 & 		&	95 $\pm$ 9	 & 	94 $\pm$ 8	 & 	454 $\pm$ 5	 & 	133 $\pm$ 6	 & 	40 $\pm$ 1	 & 	154 $\pm$ 9	 & 	1117 $\pm$ 60	 & 	144 $\pm$ 4	 \\
37742	 & 	O9.5Ibe	 & 	0.06	 & 	20.39	 $\pm$ 	0.09	 & 	15.88	 $\pm$ 	0.11	 & 	1,9	&		 & 		 & 	31 $\pm$ 4	 & 	$<$  6	 & 	2.8 $\pm$ 0.8	 & 	38 $\pm$ 5	 & 	$<$  90	 & 	6.2 $\pm$ 1.5	 \\
37903	 & 	B1.5V	 & 	0.35	 & 	21.17	 $\pm$ 	0.1	 & 	20.92	 $\pm$ 	0.06	 & 	1,10	&	46 $\pm$ 8	 & 	36 $\pm$ 5	 & 	183 $\pm$ 10	 & 	33 $\pm$ 5	 & 	11.5 $\pm$ 1.5	 & 	56.6 $\pm$ 6	 & 	503 $\pm$ 70	 & 	36 $\pm$ 4	 \\
38087	 & 	B5V	 & 	0.29	 & 				 & 	20.64	 $\pm$ 	0.07	 & 	--,10	&	$<$  25	 & 	$<$  30	 & 	162 $\pm$ 6	 & 	44 $\pm$ 6	 & 	12.6 $\pm$ 1	 & 	35 $\pm$ 6	 & 	325 $\pm$ 45	 & 	54 $\pm$ 3	 \\
38771	 & 	B0.5Ia	 & 	0.05	 & 	20.6	 $\pm$ 	0.08	 & 	15.68	 $\pm$ 	0.14	 & 	1,7	&		 & 		 & 	33 $\pm$ 5	 & 	15 $\pm$ 4	 & 	5 $\pm$ 0.8	 & 	25 $\pm$ 5	 & 	126 $\pm$ 35	 & 	7.7 $\pm$ 1.5	 \\
39777	 & 	B1.5V	 & 	0.07	 & 				 & 				 & 		&		 & 		 & 	25 $\pm$ 6	 & 	$<$  10	 & 	$<$  6	 & 	$<$  25	 & 	156 $\pm$ 45	 & 	$<$  10	 \\
40111	 & 	B0.5II	 & 	0.20	 & 	21.03	 $\pm$ 	0.09	 & 	19.73	 $\pm$ 	0.1	 & 	1,6	&	47 $\pm$ 10	 & 	44 $\pm$ 8	 & 	169 $\pm$ 6	 & 	44 $\pm$ 5	 & 	16.3 $\pm$ 1	 & 	74 $\pm$ 7	 & 	428 $\pm$ 50	 & 	37 $\pm$ 2	 \\
40893	 & 	B0IV	 & 	0.46	 & 	21.50 $\pm$ 0.10	 & 	20.58 $\pm$ 0.06	 & 		&	78 $\pm$ 6	 & 	80 $\pm$ 7	 & 	391 $\pm$ 5	 & 	109 $\pm$ 6	 & 	39 $\pm$ 1.5	 & 	203 $\pm$ 9	 & 	1030 $\pm$ 75	 & 	151 $\pm$ 4	 \\
41117	 & 	B2Iae	 & 	0.45	 & 	21.4	 $\pm$ 	0.15	 & 	20.69	 $\pm$ 	0.1	 & 	1,10	&	89 $\pm$ 6	 & 	86 $\pm$ 5	 & 	356 $\pm$ 10	 & 	148 $\pm$ 8	 & 	42 $\pm$ 1	 & 	135 $\pm$ 5	 & 	760 $\pm$ 100	 & 	154 $\pm$ 3	 \\
42087	 & 	B2.5Ibe	 & 	0.36	 & 	21.4	 $\pm$ 	0.11	 & 	20.52	 $\pm$ 	0.12	 & 	1,10	&	72 $\pm$ 5	 & 	75 $\pm$ 6	 & 	275 $\pm$ 7	 & 	99 $\pm$ 2	 & 	30 $\pm$ 1	 & 	115 $\pm$ 4	 & 	675 $\pm$ 70	 & 	115 $\pm$ 3	 \\
43247	 & 	B9II-III	 & 	0.03	 & 				 & 				 & 		&	19 $\pm$ 4	 & 	19 $\pm$ 5	 & 	64 $\pm$ 6	 & 	25 $\pm$ 5	 & 	8 $\pm$ 1	 & 	34 $\pm$ 4	 & 	190 $\pm$ 40	 & 	23 $\pm$ 2.5	 \\
43384	 & 	B3Ib	 & 	0.58	 & 	21.27 $\pm$ 0.30 	 & 	20.87	 $\pm$ 	0.14	 & 	--,10	&	109 $\pm$ 7	 & 	113 $\pm$ 4	 & 	455 $\pm$ 7	 & 	155 $\pm$ 5	 & 	48 $\pm$ 1	 & 	170 $\pm$ 6	 & 	950 $\pm$ 50	 & 	194 $\pm$ 3	 \\
46056	 & 	O8V	 & 	0.50	 & 	21.38	 $\pm$ 	0.14	 & 	20.68	 $\pm$ 	0.06	 & 	1,10	&	66 $\pm$ 7	 & 	90 $\pm$ 10	 & 	300 $\pm$ 7	 & 	135 $\pm$ 9	 & 	32 $\pm$ 1.5	 & 	151 $\pm$ 9	 & 	750 $\pm$ 60	 & 	137 $\pm$ 4	 \\
46202	 & 	O9V	 & 	0.49	 & 	21.58	 $\pm$ 	0.15	 & 	20.68	 $\pm$ 	0.07	 & 	1,10	&	55 $\pm$ 5	 & 	76 $\pm$ 10	 & 	332 $\pm$ 6	 & 	119 $\pm$ 8	 & 	35 $\pm$ 3	 & 	159 $\pm$ 10	 & 	935 $\pm$ 70	 & 	136 $\pm$ 3	 \\
46711	 & 	B3II	 & 	1.04	 & 				 & 				 & 		&	187 $\pm$ 15	 & 	173 $\pm$ 15	 & 	820 $\pm$ 10	 & 	269 $\pm$ 8	 & 	83 $\pm$ 4	 & 	287 $\pm$ 14	 & 	1500 $\pm$ 150	 & 	363 $\pm$ 5	 \\
47129	 & 	O8V+O8f	 & 	0.36	 & 	21.18	 $\pm$ 	0.11	 & 	20.55	 $\pm$ 	0.09	 & 	1,6	&	46 $\pm$ 7	 & 	30 $\pm$ 8	 & 	204 $\pm$ 5	 & 	89 $\pm$ 4	 & 	24 $\pm$ 1.5	 & 	93 $\pm$ 6	 & 	550 $\pm$ 50	 & 	89 $\pm$ 4	 \\
47839	 & 	O7Ve	 & 	0.07	 & 	20.31	 $\pm$ 	0.1	 & 	15.55	 $\pm$ 	0.09	 & 	1,7	&		 & 		 & 	30 $\pm$ 3	 & 	7 $\pm$ 2	 & 	$<$  3	 & 	12 $\pm$ 5	 & 	45 $\pm$ 15	 & 		 \\
48099	 & 	O6e	 & 	0.27	 & 	21.2	 $\pm$ 	0.12	 & 	20.29	 $\pm$ 	0.07	 & 	1,7	&	33 $\pm$ 4	 & 	43 $\pm$ 5	 & 	207 $\pm$ 7	 & 	52 $\pm$ 6	 & 	19.2 $\pm$ 0.8	 & 	87 $\pm$ 5	 & 	595 $\pm$ 50	 & 	78 $\pm$ 2	 \\
50064	 & 	B6Ia	 & 	0.85	 & 				 & 				 & 		&	190 $\pm$ 12	 & 	154 $\pm$ 12	 & 	693 $\pm$ 10	 & 	288 $\pm$ 15	 & 	72 $\pm$ 5	 & 	280 $\pm$ 12	 & 	1415 $\pm$ 150	 & 	275 $\pm$ 6	 \\
51309	 & 	B3II	 & 	0.11	 & 		 & 		& 		&	19 $\pm$ 6	 & 	17 $\pm$ 5	 & 	56 $\pm$ 5	 & 	17 $\pm$ 5	 & 	6.3 $\pm$ 1	 & 	24 $\pm$ 5	 & 	319 $\pm$ 50	 & 	11 $\pm$ 2	 \\
53367	 & 	B0IVe	 & 	0.74	 & 	21.32 $\pm$ 0.30	 & 	21.04 $\pm$ 0.05	 & 	--,10	&	39 $\pm$ 7	 & 	40 $\pm$ 6	 & 	175 $\pm$ 5	 & 	86 $\pm$ 5	 & 	23 $\pm$ 1.5	 & 	102 $\pm$ 9	 & 	542 $\pm$ 60	 & 	81 $\pm$ 3	 \\
53975	 & 	O8V	 & 	0.21	 & 	21.1	 $\pm$ 	0.08	 & 	19.23	 $\pm$ 	0.09	 & 	1,6	&	41 $\pm$ 5	 & 	36 $\pm$ 5	 & 	177 $\pm$ 5	 & 	26 $\pm$ 6	 & 	16 $\pm$ 1	 & 	92 $\pm$ 7	 & 	607 $\pm$ 60	 & 	45 $\pm$ 4	 \\
54662	 & 	O7III	 & 	0.35	 & 	21.23	 $\pm$ 	0.1	 & 	20	 $\pm$ 	0.09	 & 	1,6	&	42 $\pm$ 5	 & 	42 $\pm$ 6	 & 	217 $\pm$ 7	 & 	52 $\pm$ 8	 & 	24.5 $\pm$ 1.5	 & 	100 $\pm$ 7	 & 	595 $\pm$ 60	 & 	93 $\pm$ 4	 \\
57060	 & 	O7e+O7	 & 	0.17	 & 	20.78	 $\pm$ 	0.1	 & 	15.78	 $\pm$ 	0.1	 & 	1,7	&	23 $\pm$ 7	 & 	$<$  15	 & 	62 $\pm$ 4	 & 	$<$  25	 & 	7.8 $\pm$ 1	 & 	44 $\pm$ 9	 & 	240 $\pm$ 50	 & 	14 $\pm$ 2	 \\
57061 ($\tau$ CMa)	 & 	O9III	 & 	0.16	 & 	20.8	 $\pm$ 	0.08	 & 	15.45	 $\pm$ 	0.13	 & 	1,7	&	$<$  20	 & 	16 $\pm$ 5	 & 	63 $\pm$ 4	 & 	$<$  20	 & 	6.6 $\pm$ 0.8	 & 	56 $\pm$ 5	 & 	272 $\pm$ 45	 & 	11 $\pm$ 2	 \\
90994	 & 	B6V	 & 	0.00	 & 				 & 				 & 		&		 & 		 & 	14 $\pm$ 3	 & 	$<$  12	 & 	$<$  2.5	 & 	$<$  15	 & 	$<$  150	 & 	$<$  6	 \\
91316	 & 	B1Ib	 & 	0.05	 & 	20.44	 $\pm$ 	0.09	 & 	15.58	 $\pm$ 	0.08	 & 	1,6	&	$<$  20	 & 	12 $\pm$ 3	 & 	38 $\pm$ 3	 & 	17 $\pm$ 3	 & 	3.6 $\pm$ 0.6	 & 	20 $\pm$ 5	 & 	75 $\pm$ 20	 & 	10 $\pm$ 3	 \\
97991	 & 	B1.5V	 & 	0.04	 & 	20.54	 $\pm$ 	0.08	 & 	15.99	 $\pm$ 	0.2	 & 	1,11	&		 & 		 & 	32 $\pm$ 6	 & 	16 $\pm$ 5	 & 	$<$  4	 & 	$<$  30	 & 	135 $\pm$ 45	 & 	9.5 $\pm$ 2.5	 \\
143018	 & 	B1V+B2V	 & 	0.05	 & 	20.66	 $\pm$ 	0.1	 & 	19.32	 $\pm$ 	0.1	 & 	1,6	&	$<$  20	 & 	$<$  18	 & 	39 $\pm$ 4	 & 	7 $\pm$ 2	 & 	2.5 $\pm$ 0.8	 & 	14 $\pm$ 5	 & 	145 $\pm$ 40	 & 	10 $\pm$ 2	 \\
143275	 & 	B0.3IV	 & 	0.17	 & 	21.01	 $\pm$ 	0.08	 & 	19.42	 $\pm$ 	0.1	 & 	1,6	&	27 $\pm$ 5	 & 	19 $\pm$ 6	 & 	82 $\pm$ 5	 & 	26 $\pm$ 4	 & 	7.8 $\pm$ 0.8	 & 	30 $\pm$ 4	 & 	250 $\pm$ 25	 & 	23 $\pm$ 3	 \\
144217 ($\beta^1$ Sco AB)	 & 	B1V	 & 	0.19	 & 	21.03	 $\pm$ 	0.08	 & 	19.83	 $\pm$ 	0.04	 & 	1,6	&	28 $\pm$ 9	 & 	39 $\pm$ 5	 & 	171 $\pm$ 5	 & 	34 $\pm$ 4	 & 	13.2 $\pm$ 0.8	 & 	57.6 $\pm$ 5	 & 	397 $\pm$ 45	 & 	42 $\pm$ 3	 \\
144218 ($\beta^2$ Sco)	 & 	B2V	 & 	0.22	 & 				 & 				 & 		&	41 $\pm$ 9	 & 	44 $\pm$ 6	 & 	191 $\pm$ 5	 & 	36 $\pm$ 5	 & 	13 $\pm$ 1	 & 	62 $\pm$ 7	 & 	404 $\pm$ 45	 & 	47 $\pm$ 4	 \\
144470 ($\omega^1$ Sco)	 & 	B1V	 & 	0.22	 & 	21.18	 $\pm$ 	0.08	 & 	20.06	 $\pm$ 	0.06	 & 	5,6	&	37 $\pm$ 8	 & 	37 $\pm$ 6	 & 	192 $\pm$ 5	 & 	40 $\pm$ 4	 & 	17 $\pm$ 1	 & 	58 $\pm$ 6	 & 	403 $\pm$ 40	 & 	63 $\pm$ 3	 \\
145502 ($\nu$ Sco AB)	 & 	B3V	 & 	0.24	 & 	21.2	 $\pm$ 	0.12	 & 	19.89	 $\pm$ 	0.07	 & 	1,6	&	26 $\pm$ 8	 & 	37 $\pm$ 7	 & 	187 $\pm$ 5	 & 	49 $\pm$ 5	 & 	16 $\pm$ 1	 & 	55 $\pm$ 6	 & 	421 $\pm$ 40	 & 	63 $\pm$ 3	 \\
147165 ($\sigma$ Sco)	 & 	B2III+O9V	 & 	0.41	 & 	21.38	 $\pm$ 	0.08	 & 	19.79	 $\pm$ 	0.07	 & 	1,6	&	51 $\pm$ 5	 & 	64 $\pm$ 5	 & 	254 $\pm$ 5	 & 	54 $\pm$ 3	 & 	15.2 $\pm$ 0.7	 & 	75 $\pm$ 7	 & 	498 $\pm$ 50	 & 	63 $\pm$ 3	 \\
147888	 & 	B5V	 & 	0.47	 & 	21.44 $\pm$ 0.30	 & 	20.47	 $\pm$ 	0.05	 & 	--,10	&	45 $\pm$ 5	 & 	54 $\pm$ 4	 & 	252 $\pm$ 12	 & 	60 $\pm$ 5	 & 	19 $\pm$ 1	 & 	56 $\pm$ 5	 & 	390 $\pm$ 60	 & 	82 $\pm$ 2	 \\
147889	 & 	B2V	 & 	1.07	 & 				 & 				 & 		&	75 $\pm$ 6	 & 	85 $\pm$ 6	 & 	377 $\pm$ 8	 & 	163 $\pm$ 5	 & 	46 $\pm$ 2	 & 	95 $\pm$ 7	 & 	530 $\pm$ 50	 & 	180 $\pm$ 5	 \\
147933 ($\rho$ Oph A)	 & 	B2IV	 & 	0.48	 & 	21.63	 $\pm$ 	0.09	 & 	20.57	 $\pm$ 	0.07	 & 	1,6	&	55 $\pm$ 6	 & 	44 $\pm$ 8	 & 	222 $\pm$ 10	 & 	71 $\pm$ 6	 & 	17.1 $\pm$ 1	 & 	50 $\pm$ 7	 & 	426 $\pm$ 80	 & 	68 $\pm$ 5	 \\
148184	 & 	B2IVpe	 & 	0.52	 & 				 & 	20.63	 $\pm$ 	0.09	 & 	--,6	&		 & 		 & 	102 $\pm$ 5	 & 	64 $\pm$ 4	 & 	14 $\pm$ 1	 & 	55 $\pm$ 5	 & 	327 $\pm$ 45	 & 	40 $\pm$ 3	 \\
148605	 & 	B2V	 & 	0.13	 & 				 & 	18.74	 $\pm$ 	0.09	 & 	--,6	&	$<$  15	 & 	$<$  15	 & 	51 $\pm$ 4	 & 	12 $\pm$ 4	 & 	3.8 $\pm$ 1	 & 	15 $\pm$ 5	 & 	173 $\pm$ 45	 & 	9.4 $\pm$ 2	 \\
149404	 & 	O9Iae	 & 	0.68	 & 	21.4	 $\pm$ 	0.14	 & 				 & 	1,10	&	142 $\pm$ 10	 & 	94 $\pm$ 6	 & 	436 $\pm$ 8	 & 	112 $\pm$ 10	 & 	42 $\pm$ 1.5	 & 	158 $\pm$ 8	 & 	900 $\pm$ 60	 & 	170 $\pm$ 3	 \\
149881	 & 	B0.5III	 & 	0.10	 & 	20.57	 $\pm$ 	0.08	 & 	19.09	 $\pm$ 	0.1	 & 	1,6	&	$<$  15	 & 	$<$  18	 & 	44 $\pm$ 4	 & 	12 $\pm$ 3	 & 	4.8 $\pm$ 0.8	 & 	25 $\pm$ 5	 & 	147 $\pm$ 40	 & 	9 $\pm$ 1.5	 \\
149757	 & 	O9.5V	 & 	0.32	 & 	20.69	 $\pm$ 	0.1	 & 	20.66	 $\pm$ 	0.04	 & 	1,6	&	11 $\pm$ 3	 & 	$<$  18	 & 	83 $\pm$ 7	 & 	38 $\pm$ 4	 & 	10 $\pm$ 1	 & 	36 $\pm$ 5	 & 	175 $\pm$ 35	 & 	41 $\pm$ 3	 \\
157857	 & 	O7e	 & 	0.51	 & 				 & 				 & 		&	40 $\pm$ 5	 & 	43 $\pm$ 5	 & 	265 $\pm$ 5	 & 	72 $\pm$ 5	 & 	28 $\pm$ 1	 & 	118 $\pm$ 8	 & 	905 $\pm$ 60	 & 	114 $\pm$ 5	 \\
159975	 & 	B8II-IIIp	 & 	0.19	 & 				 & 				 & 		&	32 $\pm$ 5	 & 	33 $\pm$ 4	 & 	189 $\pm$ 5	 & 	72 $\pm$ 4	 & 	17 $\pm$ 1	 & 	60 $\pm$ 5	 & 	500 $\pm$ 50	 & 	64 $\pm$ 3	 \\
162978	 & 	O8III	 & 	0.35	 & 	21.28	 $\pm$ 	0.08	 & 				 & 	1,--	&	42 $\pm$ 7	 & 	44 $\pm$ 8	 & 	211 $\pm$ 14	 & 	58 $\pm$ 9	 & 	21.5 $\pm$ 2	 & 	88 $\pm$ 11	 & 	558 $\pm$ 60	 & 	64 $\pm$ 5	 \\
164353	 & 	B5Ib	 & 	0.11	 & 	21			 & 	20.26	 $\pm$ 	0.14	 & 	5,6	&	24 $\pm$ 5	 & 	24 $\pm$ 4	 & 	124 $\pm$ 4	 & 	38 $\pm$ 4	 & 	12.9 $\pm$ 0.7	 & 	46 $\pm$ 5	 & 	378 $\pm$ 45	 & 	53 $\pm$ 3	 \\
164740 (Herschel 36)	 & 	O7.5V	 & 	0.87	 & 				 & 	20.19	 $\pm$ 	0.12	 & 	--,10	&	80 $\pm$ 9	 & 	119 $\pm$ 18	 & 	463 $\pm$ 8	 & 	102 $\pm$ 7	 & 	28 $\pm$ 2	 & 	142 $\pm$ 9	 & 	880 $\pm$ 75	 & 	132 $\pm$ 6	 \\
166734	 & 	O8e	 & 	1.39	 & 				 & 				 & 		&	216 $\pm$ 10	 & 	168 $\pm$ 9	 & 	727 $\pm$ 8	 & 	322 $\pm$ 15	 & 	93 $\pm$ 3	 & 	321 $\pm$ 12	 & 	1560 $\pm$ 200	 & 	401 $\pm$ 5	 \\
166937 ($\mu$ Sgr)	 & 	B8Iape	 & 	0.25	 & 				 & 				 & 		&	45.5 $\pm$ 3	 & 	63 $\pm$ 5	 & 	283 $\pm$ 4	 & 	100 $\pm$ 5	 & 	25.7 $\pm$ 1	 & 	90 $\pm$ 6	 & 	785 $\pm$ 70	 & 	90 $\pm$ 3	 \\
167971	 & 	O8e	 & 	1.08	 & 	21.6	 $\pm$ 	0.3	 & 	20.85	 $\pm$ 	0.12	 & 	8,8	&	116 $\pm$ 10	 & 	131 $\pm$ 5	 & 	512 $\pm$ 9	 & 	208 $\pm$ 6	 & 	58 $\pm$ 2	 & 	241 $\pm$ 10	 & 	1450 $\pm$ 200	 & 	219 $\pm$ 3	 \\
168076	 & 	O5f	 & 	0.78	 & 	21.65	 $\pm$ 	0.23	 & 	20.68	 $\pm$ 	0.08	 & 	1,8	&	110 $\pm$ 15	 & 	118 $\pm$ 12	 & 	541 $\pm$ 10	 & 	250 $\pm$ 8	 & 	59.5 $\pm$ 3	 & 	219 $\pm$ 15	 & 	1090 $\pm$ 150	 & 	221 $\pm$ 6	 \\
169454	 & 	B1.5Ia	 & 	1.12	 & 				 & 				 & 		&	128 $\pm$ 10	 & 	118 $\pm$ 5	 & 	510 $\pm$ 7	 & 	213 $\pm$ 5	 & 	57 $\pm$ 1	 & 	219 $\pm$ 10	 & 	1580 $\pm$ 170	 & 	205 $\pm$ 5	 \\
170740	 & 	B2V	 & 	0.48	 & 	21.04	 $\pm$ 	0.15	 & 	20.86	 $\pm$ 	0.08	 & 	1,8	&	66 $\pm$ 7	 & 	63 $\pm$ 5	 & 	255 $\pm$ 7	 & 	92 $\pm$ 5	 & 	26.6 $\pm$ 0.8	 & 	76 $\pm$ 4	 & 	595 $\pm$ 60	 & 	125 $\pm$ 4	 \\
172028	 & 	B2V	 & 	0.79	 & 				 & 				 & 		&	50 $\pm$ 10	 & 	53 $\pm$ 5	 & 	256 $\pm$ 8	 & 	217 $\pm$ 5	 & 	37 $\pm$ 1	 & 	94 $\pm$ 5	 & 	450 $\pm$ 60	 & 	137 $\pm$ 3	 \\
175156	 & 	B5II	 & 	0.31	 & 				 & 				 & 		&	24 $\pm$ 4	 & 	36 $\pm$ 4	 & 	151 $\pm$ 4	 & 	85 $\pm$ 5	 & 	18 $\pm$ 0.8	 & 	59 $\pm$ 6	 & 	418 $\pm$ 55	 & 	69 $\pm$ 4	 \\
179406	 & 	B3V	 & 	0.33	 & 	21.23 $\pm$ 0.15	 & 	20.73	 $\pm$ 	0.07	 & 	--,10	&	25 $\pm$ 5	 & 	29 $\pm$ 6	 & 	172 $\pm$ 5	 & 	76 $\pm$ 3	 & 	19.8 $\pm$ 0.7	 & 	44 $\pm$ 3	 & 	430 $\pm$ 60	 & 	98 $\pm$ 2	 \\
183143	 & 	B7Iae	 & 	1.27	 & 				 & 				 & 		&	225 $\pm$ 14	 & 	172 $\pm$ 7	 & 	761 $\pm$ 6	 & 	257 $\pm$ 8	 & 	89 $\pm$ 2	 & 	340 $\pm$ 11	 & 	1910 $\pm$ 30	 & 	332 $\pm$ 4	 \\
185418	 & 	B0.5V	 & 	0.50	 & 	21.11	 $\pm$ 	0.15	 & 	20.76	 $\pm$ 	0.05	 & 	8,8	&	57 $\pm$ 6	 & 	57 $\pm$ 3	 & 	273 $\pm$ 5	 & 	105 $\pm$ 5	 & 	35 $\pm$ 1	 & 	111 $\pm$ 6	 & 	640 $\pm$ 50	 & 	164 $\pm$ 4	 \\
186994	 & 	B0III	 & 	0.17	 & 	20.9	 $\pm$ 	0.15	 & 	19.59	 $\pm$ 	0.04	 & 	5,10	&	40 $\pm$ 6	 & 	$<$  30	 & 	101 $\pm$ 5	 & 	23 $\pm$ 4	 & 	11 $\pm$ 1	 & 	60 $\pm$ 6	 & 	296 $\pm$ 40	 & 	17 $\pm$ 3	 \\
192639	 & 	O8e	 & 	0.66	 & 	21.32	 $\pm$ 	0.12	 & 	20.69	 $\pm$ 	0.05	 & 	1,8	&	75 $\pm$ 6	 & 	81 $\pm$ 5	 & 	324 $\pm$ 5	 & 	79 $\pm$ 7	 & 	39 $\pm$ 1	 & 	151 $\pm$ 5	 & 	817 $\pm$ 50	 & 	150 $\pm$ 3	 \\
194839	 & 	B0.5Ia	 & 	1.18	 & 				 & 				 & 		&	216 $\pm$ 16	 & 	153 $\pm$ 20	 & 	585 $\pm$ 6	 & 	186 $\pm$ 7	 & 	56 $\pm$ 2	 & 	289 $\pm$ 7	 & 	1690 $\pm$ 100	 & 	173 $\pm$ 4	 \\
Cyg OB2 5	 & 	O7f	 & 	1.99	 & 				 & 				 & 		&	251 $\pm$ 15	 & 	195 $\pm$ 20	 & 	774 $\pm$ 8	 & 	239 $\pm$ 12	 & 	83 $\pm$ 1	 & 	363 $\pm$ 10	 & 	1990 $\pm$ 200	 & 	312 $\pm$ 8	 \\
Cyg OB2 12	 & 	B5Ie	 & 	3.31	 & 				 & 				 & 		&	225 $\pm$ 30	 & 	214 $\pm$ 15	 & 	850 $\pm$ 20	 & 	381 $\pm$ 15	 & 	103 $\pm$ 3	 & 	395 $\pm$ 9	 & 	2215 $\pm$ 200	 & 	377 $\pm$ 6	 \\
198478	 & 	B3Iae	 & 	0.54	 & 				 & 				 & 		&	90 $\pm$ 6	 & 	72 $\pm$ 4	 & 	332 $\pm$ 5	 & 	112 $\pm$ 4	 & 	33.1 $\pm$ 1.5	 & 	130 $\pm$ 7	 & 	919 $\pm$ 60	 & 	139 $\pm$ 3	 \\
199579	 & 	O6Ve	 & 	0.37	 & 	21.04	 $\pm$ 	0.11	 & 	20.53	 $\pm$ 	0.04	 & 	1,8	&	32 $\pm$ 4	 & 	21 $\pm$ 3	 & 	128 $\pm$ 5	 & 	50 $\pm$ 4	 & 	15.5 $\pm$ 1	 & 	53 $\pm$ 2	 & 	315 $\pm$ 50	 & 	63 $\pm$ 2	 \\
199892	 & 	B7III	 & 	0.04	 & 				 & 				 & 		&		 & 		 & 	29 $\pm$ 9	 & 	$<$  9	 & 	$<$  3	 & 	16 $\pm$ 5	 & 	150 $\pm$ 45	 & 	$<$  6	 \\
201345	 & 	O9.5V	 & 	0.18	 & 				 & 				 & 		&	46 $\pm$ 7	 & 	15 $\pm$ 5	 & 	100 $\pm$ 6	 & 	29 $\pm$ 5	 & 	8.9 $\pm$ 1	 & 	52 $\pm$ 6	 & 	385 $\pm$ 55	 & 	21 $\pm$ 3.5	 \\
202850	 & 	B9Iab	 & 	0.12	 & 				 & 				 & 		&	37 $\pm$ 7	 & 	47 $\pm$ 7	 & 	173 $\pm$ 6	 & 	55 $\pm$ 6	 & 	15 $\pm$ 1	 & 	82 $\pm$ 10	 & 	576 $\pm$ 50	 & 	42 $\pm$ 3	 \\
203938	 & 	B0.5IV	 & 	0.74	 & 	21.48	 $\pm$ 	0.15	 & 	21	 $\pm$ 	0.06	 & 	8,8	&	78 $\pm$ 6	 & 	68 $\pm$ 4	 & 	356 $\pm$ 5	 & 	152 $\pm$ 5	 & 	42 $\pm$ 1	 & 	151 $\pm$ 5	 & 	936 $\pm$ 60	 & 	146 $\pm$ 3	 \\
204172	 & 	B0Ib	 & 	0.16	 & 	21	 $\pm$ 	0.11	 & 	19.6	 $\pm$ 	0.09	 & 	5,6	&	43 $\pm$ 5	 & 	19 $\pm$ 5	 & 	120 $\pm$ 4	 & 	31.6 $\pm$ 3	 & 	12.1 $\pm$ 1	 & 	57.6 $\pm$ 4	 & 	322 $\pm$ 60	 & 	33 $\pm$ 2	 \\
204827	 & 	B0V	 & 	1.11	 & 				 & 				 & 		&	68 $\pm$ 4	 & 	58 $\pm$ 3	 & 	257 $\pm$ 4	 & 	199 $\pm$ 3	 & 	41.5 $\pm$ 1	 & 	116 $\pm$ 4	 & 	518 $\pm$ 60	 & 	171 $\pm$ 3	 \\
206165	 & 	B2Ib	 & 	0.47	 & 				 & 				 & 		&	60.5 $\pm$ 4	 & 	58 $\pm$ 5	 & 	231 $\pm$ 7	 & 	106 $\pm$ 5	 & 	26 $\pm$ 1	 & 	86 $\pm$ 6	 & 	486 $\pm$ 60	 & 	111 $\pm$ 3	 \\
206267	 & 	O6f	 & 	0.53	 & 	21.3	 $\pm$ 	0.15	 & 	20.86	 $\pm$ 	0.04	 & 	8,8	&	53 $\pm$ 7	 & 	59 $\pm$ 4	 & 	242 $\pm$ 7	 & 	102 $\pm$ 5	 & 	29 $\pm$ 1	 & 	103 $\pm$ 5	 & 	544 $\pm$ 45	 & 	126 $\pm$ 3	 \\
206773	 & 	B0Vpe	 & 	0.54	 & 				 & 				 & 		&	34 $\pm$ 7	 & 	20 $\pm$ 6	 & 	193 $\pm$ 6	 & 	71 $\pm$ 6	 & 	21.5 $\pm$ 1	 & 	67 $\pm$ 7	 & 	461 $\pm$ 60	 & 	90 $\pm$ 4	 \\
207198	 & 	O9IIe	 & 	0.62	 & 	21.34	 $\pm$ 	0.17	 & 	20.83	 $\pm$ 	0.04	 & 	1,8	&	45 $\pm$ 6	 & 	56 $\pm$ 5	 & 	262 $\pm$ 6	 & 	144 $\pm$ 3	 & 	30 $\pm$ 1	 & 	111 $\pm$ 5	 & 	543 $\pm$ 40	 & 	125 $\pm$ 3	 \\
208440	 & 	B1V	 & 	0.33	 & 				 & 				 & 		&	42 $\pm$ 7	 & 	50 $\pm$ 7	 & 	213 $\pm$ 7	 & 	92 $\pm$ 6	 & 	21 $\pm$ 1.5	 & 	100 $\pm$ 7	 & 	597 $\pm$ 65	 & 	98 $\pm$ 4	 \\
208501	 & 	B8Ib	 & 	0.75	 & 				 & 				 & 		&	60 $\pm$ 7	 & 	52 $\pm$ 6	 & 	255 $\pm$ 6	 & 	128 $\pm$ 6	 & 	36 $\pm$ 1	 & 	110 $\pm$ 9	 & 	666 $\pm$ 65	 & 	124 $\pm$ 4	 \\
209008	 & 	B3III	 & 	0.08	 & 				 & 				 & 		&	$<$  25	 & 	$<$  20	 & 	46 $\pm$ 6	 & 	14 $\pm$ 4	 & 	5 $\pm$ 1	 & 	15 $\pm$ 5	 & 	227 $\pm$ 40	 & 	9.5 $\pm$ 2	 \\
209975	 & 	O9Ib	 & 	0.36	 & 	21.17	 $\pm$ 	0.09	 & 	20.08	 $\pm$ 	0.09	 & 	1,6	&	64 $\pm$ 10	 & 	43 $\pm$ 8	 & 	258 $\pm$ 5	 & 	91 $\pm$ 5	 & 	29 $\pm$ 1.5	 & 	96 $\pm$ 8	 & 	520 $\pm$ 60	 & 	115 $\pm$ 4	 \\
210121	 & 	B3V	 & 	0.40	 & 	20.63	 $\pm$ 	0.15	 & 	20.75	 $\pm$ 	0.12	 & 	8,8	&	15.5 $\pm$ 3.5	 & 	$<$  20	 & 	70 $\pm$ 7	 & 	46 $\pm$ 9	 & 	9.4 $\pm$ 0.7	 & 	27.5 $\pm$ 4	 & 	146 $\pm$ 50	 & 	25 $\pm$ 2	 \\
210839	 & 	O6If	 & 	0.57	 & 	21.15	 $\pm$ 	0.12	 & 	20.84	 $\pm$ 	0.04	 & 	1,8	&	52 $\pm$ 5	 & 	65 $\pm$ 5	 & 	261 $\pm$ 5	 & 	72 $\pm$ 6	 & 	31 $\pm$ 1	 & 	106 $\pm$ 5	 & 	551 $\pm$ 45	 & 	150 $\pm$ 3	 \\
212120	 & 	B6V	 & 	0.04	 & 				 & 				 & 		&		 & 		 & 	57 $\pm$ 7	 & 	11 $\pm$ 2.5	 & 	4.5 $\pm$ 0.8	 & 	12 $\pm$ 4	 & 	128 $\pm$ 35	 & 	9.5 $\pm$ 2	 \\
212791	 & 	B3V	 & 	0.17	 & 				 & 				 & 		&	23 $\pm$ 7	 & 	23 $\pm$ 6	 & 	123 $\pm$ 4	 & 	33 $\pm$ 5	 & 	12.8 $\pm$ 0.8	 & 	57 $\pm$ 5	 & 	433 $\pm$ 55	 & 	37 $\pm$ 3	 \\
214080	 & 	B1Ib	 & 	0.06	 & 	20.58	 $\pm$ 	0.1	 & 	18.35	 $\pm$ 	0.1	 & 	1,9	&	22 $\pm$ 7	 & 	14 $\pm$ 4	 & 	49 $\pm$ 5	 & 	23 $\pm$ 5	 & 	$<$  4.5	 & 	29 $\pm$ 8	 & 	294 $\pm$ 60	 & 	5 $\pm$ 1.5	 \\
214680	 & 	O9V	 & 	0.11	 & 	20.69	 $\pm$ 	0.14	 & 	19.22	 $\pm$ 	0.06	 & 	1,6	&	$<$  8	 & 	$<$  27	 & 	74 $\pm$ 5	 & 	25 $\pm$ 3	 & 	5.9 $\pm$ 0.7	 & 	20 $\pm$ 5	 & 	237 $\pm$ 50	 & 	15 $\pm$ 1	 \\
214930	 & 	B2IV	 & 	0.10	 & 				 & 				 & 		&	$<$  15	 & 	$<$  12	 & 	53 $\pm$ 4	 & 	18 $\pm$ 3	 & 	7.5 $\pm$ 1.5	 & 	20 $\pm$ 5	 & 	161 $\pm$ 40	 & 	16 $\pm$ 3	 \\
215733	 & 	B1II	 & 	0.11	 & 	20.75	 $\pm$ 	0.09	 & 	19.45	 $\pm$ 	0.1	 & 	1,9	&	33 $\pm$ 6	 & 	24 $\pm$ 4	 & 	82 $\pm$ 5	 & 	32 $\pm$ 4	 & 	8.6 $\pm$ 1	 & 	40 $\pm$ 6	 & 	246 $\pm$ 45	 & 	21.5 $\pm$ 2	 \\
218376	 & 	B0.5IV	 & 	0.25	 & 	20.91	 $\pm$ 	0.09	 & 	20.15	 $\pm$ 	0.09	 & 	1,6	&	36 $\pm$ 6	 & 	45 $\pm$ 5	 & 	146 $\pm$ 8	 & 	61.7 $\pm$ 6	 & 	14.9 $\pm$ 1	 & 	55 $\pm$ 6	 & 	365 $\pm$ 45	 & 	68 $\pm$ 3	 \\
219188	 & 	B0.5II	 & 	0.13	 & 	20.75	 $\pm$ 	0.09	 & 	19.38	 $\pm$ 	0.12	 & 	1,6	&	16 $\pm$ 5	 & 	14 $\pm$ 4.5	 & 	70 $\pm$ 4	 & 	27 $\pm$ 4	 & 	7.2 $\pm$ 0.8	 & 	44 $\pm$ 5	 & 	370 $\pm$ 55	 & 	19 $\pm$ 3	 \\
BD+63 1964	 & 	B0II	 & 	1.00	 & 				 & 				 & 		&	210 $\pm$ 12	 & 	195 $\pm$ 20	 & 	729 $\pm$ 10	 & 	295 $\pm$ 9	 & 	92 $\pm$ 2	 & 	313 $\pm$ 15	 & 	1380 $\pm$ 200	 & 	330 $\pm$ 6	 \\
223385 (6 Cas)	 & 	A3Iae	 & 	0.67	 & 				 & 				 & 		&	85 $\pm$ 8	 & 	166 $\pm$ 15	 & 	582 $\pm$ 6	 & 	167 $\pm$ 6	 & 	35 $\pm$ 3	 & 	230 $\pm$ 8	 & 	1360 $\pm$ 70	 & 	193 $\pm$ 5	 \\
224572	 & 	B1V	 & 	0.19	 & 	20.79	 $\pm$ 	0.08	 & 	20.23	 $\pm$ 	0.09	 & 	1,6	&	$<$  20	 & 	15 $\pm$ 5	 & 	78 $\pm$ 4	 & 	31 $\pm$ 4	 & 	9 $\pm$ 1	 & 	27 $\pm$ 5	 & 	276 $\pm$ 40	 & 	30 $\pm$ 3	 \\
229059	 & 	B1.5Iap	 & 	1.71	 & 				 & 				 & 		&	116 $\pm$ 12	 & 	96 $\pm$ 5	 & 	457 $\pm$ 7	 & 	163 $\pm$ 3	 & 	61 $\pm$ 1	 & 	212 $\pm$ 10	 & 	1090 $\pm$ 150	 & 	241 $\pm$ 5	 \\

\enddata
\tablenotetext{a}{The calculated values of \ebv are based on the intrinsic colors from Johnson (1963).}
\tablenotetext{b}{The references for \NH\ and \Nhtwo, respectively. References -- (1) Diplas \& Savage (1994), Table 1;  (2) Diplas \& Savage (1994), Table 2; (3) Shull \& Van Steenberg (1985), Table 1; (4) Shull \& Van Steenberg (1985), Table 2; (5) Bohlin, Savage, \& Drake (1978);
(6) Savage et al. (1977); (7) Spitzer, Cochran, \& Hirshfeld (1974); (8) Rachford et al. (2002); (9) B. Rachford, unpublished; (10) Rachford et al. (2009); (11) K. Gillmon, private communication; (12) Jenkins et al. (1999); (13) Jenkins et al. (2000)}
\end{deluxetable}
\clearpage
\end{landscape}


\tabletypesize{\scriptsize}
\begin{deluxetable}{lcccccc}
\tablenum{2}
\tablecolumns{7}
\tablewidth{0pt}
\tablecaption{DIB Correlation Data with log[\NH]\tablenotemark{a,b}}
\tablehead{
\colhead{DIB (FWHM\tablenotemark{c})} &
\colhead{Correlation Coefficient} &
\colhead{Reduced $\chi^2$} &
\colhead{Number of sight lines} &
\colhead{Correlation Coefficient\tablenotemark{d}} &
\colhead{a\tablenotemark{e}} &
\colhead{b\tablenotemark{e}}
}
\startdata
5780.5 (2.11) & 0.94  & 1.209 & 74 & 0.90 & $19.00 \pm 0.08$ & $0.94 \pm 0.04$ \\
6283.8 (4.77) & 0.89  & 1.250 & 71 & 0.87 & $17.65 \pm 0.20$ & $1.30 \pm 0.07$ \\
6204.5 (4.87) & 0.89  & 1.559 & 69 & 0.84 & $ 19.02 \pm 0.11$ & $1.12 \pm 0.06$ \\
6196.0 (0.42) & 0.89  & 2.035 & 68 & 0.79 & $ 19.90 \pm 0.06$ & $0.95 \pm 0.05$ \\
6613.6 (0.93) & 0.87  & 2.794 & 70 & 0.77 & $19.89 \pm 0.06$ & $0.67 \pm 0.03$ \\
5705.1 (2.58) & 0.83  & 1.278 & 52 & 0.73 & $19.38 \pm 0.16$ & $1.08 \pm 0.09$ \\
5797.1 (0.77) & 0.82  & 3.269 & 65 & 0.72 & $19.59 \pm 0.09$ & $0.85 \pm 0.05$ \\
5487.7 (5.20) & 0.78  & 1.516 & 55 & 0.60 & $19.28 \pm 0.16$ & $1.13 \pm 0.09$ \\
\enddata
\tablenotetext{a}{All entries computed using logarithmic values of DIB equivalent widths}
\tablenotetext{b}{Data in columns 2, 3, 4, 6, and 7 exclude $\rho$ Oph A, $\theta^1$ Ori C, and HD 37061}
\tablenotetext{c}{FWHM from Paper II. All wavelengths are in units of \AA}
\tablenotetext{d}{Includes all sight lines}
\tablenotetext{e}{Coefficients for log[\NH]  = a + b$\times$log$[W_{\lambda}$(DIB)]}
\end{deluxetable}


\begin{deluxetable}{lcccc}
\tablenum{3}
\tablecolumns{5}
\tablewidth{0pt}
\tablecaption{DIB Correlation Data with log[\Nhtwo]\tablenotemark{a,b}}
\tablehead{
\colhead{DIB} &
\colhead{Correlation Coefficient} &
\colhead{Reduced $\chi^2$} &
\colhead{Number of sight lines} &
\colhead{Correlation Coefficient\tablenotemark{c}}
}
\startdata
5780.5 & 0.65  & 39.60 & 64 & 0.65 \\
6283.8 & 0.46  & 9.99 & 63 & 0.46 \\
6204.5 & 0.60  & 24.61 & 63 & 0.60 \\
6196.0 & 0.74  & 21.10 & 63 & 0.74 \\
6613.6 & 0.80  & 21.82 & 64 & 0.80 \\
5705.1 & 0.56  & 10.62 & 49 & 0.56 \\
5797.1 & 0.79  & 14.83 & 63 & 0.79 \\
5487.7 & 0.47  & 10.42 & 52 & 0.47 \\
\enddata
\tablenotetext{a}{All entries computed using logarithmic values of DIB equivalent widths and only for sight
lines with log[\Nhtwo] $> 18$}
\tablenotetext{b}{Data in columns 2, 3, and 4 exclude $\rho$ Oph A, $\theta^1$ Ori C, and HD 37061}
\tablenotetext{c}{Includes all sight lines}
\end{deluxetable}


\begin{landscape}
\begin{deluxetable}{lcccccc}
\tablenum{4}
\tablecolumns{7}
\tablewidth{0pt}
\tablecaption{DIB Correlation Data with \ebv\tablenotemark{a,b}}
\tablehead{
\colhead{DIB} &
\colhead{Correlation Coefficient} &
\colhead{Reduced $\chi^2$} &
\colhead{Number of sight lines} &
\colhead{Correlation Coefficient\tablenotemark{c}} &
\colhead{a\tablenotemark{d}} &
\colhead{b\tablenotemark{d}}
}
\startdata
5780.5 & 0.82  & 48.44 & 133 & 0.82 & $(-8.36 \pm 3.48) \times 10^{-3}$ & $(1.98 \pm 0.01) \times 10^{-3}$ \\
6283.8 & 0.82  & 16.60 & 127 & 0.82 & $(-7.71 \pm 0.78) \times 10^{-2}$ & $(9.57 \pm 0.17) \times 10^{-4}$ \\
6204.5 & 0.83  & 26.71 & 119 & 0.83  & $(-7.22 \pm 0.67) \times 10^{-2}$ & $(5.99 \pm 0.08) \times 10^{-3}$ \\
6196.0 & 0.85  & 25.59 & 117 & 0.85  & $(-5.07 \pm 0.56) \times 10^{-2}$ & $(2.11 \pm 0.02) \times 10^{-2}$ \\
6613.6 & 0.84  & 45.27 & 120 & 0.83  & $(1.96\pm 0.37) \times 10^{-2}$ & $(4.63 \pm 0.04) \times 10^{-3}$ \\
5705.1 & 0.80  & 13.55 &   91 & 0.80  & $(-1.74 \pm 0.16) \times 10^{-1}$ & $(1.20 \pm 0.03) \times 10^{-2}$ \\
5797.1 & 0.84  & 25.16 & 113 & 0.84 & $(-2.86 \pm 0.57) \times 10^{-2}$ & $(5.74 \pm 0.06) \times 10^{-3}$  \\
5487.7 & 0.80  & 12.84 &   93 & 0.79  & $(-6.41 \pm 1.31) \times 10^{-2}$ & $(9.67 \pm 0.25) \times 10^{-3}$ \\
\enddata
\tablenotetext{a}{All entries computed using linear values of DIB equivalent widths}
\tablenotetext{b}{Data in columns 2, 3, 4, 6, and 7 exclude $\rho$ Oph A, $\theta^1$ Ori C, and HD 37061}
\tablenotetext{c}{Includes all sight lines}
\tablenotetext{d}{Coefficients for \ebv = a + b$\times W_\lambda$(5780.5)}
\end{deluxetable}
\clearpage
\end{landscape}



\begin{deluxetable}{l|cccccccc}
\tablenum{5}
\tablecolumns{9}
\tablewidth{0pt}
\tablecaption{DIB$-$DIB Correlation Coefficients\tablenotemark{a,b}}
\tablehead{
\colhead{ } &
\colhead{5780.5} &
\colhead{6204.5} &
\colhead{6196.0} &
\colhead{6283.8} &
\colhead{6613.6} &
\colhead{5705.1} &
\colhead{5797.1} &
\colhead{5487.7}
}
\startdata
5780.5 & 1 & 0.97 & 0.97 & 0.96 & 0.96 & 0.98 & 0.93 & 0.95\\
6204.5 & 0.97 & 1 & 0.96 & 0.98 & 0.94 & 0.96 & 0.91 & 0.95\\
6196.0 & 0.97 & 0.96 & 1 & 0.93 & 0.99 & 0.93 & 0.96 & 0.94\\
6283.8 & 0.96 & 0.98 & 0.93 & 1 & 0.91 & 0.94 & 0.86 & 0.92\\
6613.6 & 0.96 & 0.94 & 0.99 & 0.91 & 1 & 0.93 & 0.95 & 0.91\\
5705.1 & 0.98 & 0.96 & 0.93 & 0.94 & 0.93 & 1 & 0.90 & 0.93\\
5797.1 & 0.93 & 0.91 & 0.96 & 0.86 & 0.95 & 0.90 & 1 & 0.87\\
5487.7 & 0.95 & 0.95 & 0.94 & 0.92 & 0.91 & 0.93 & 0.87 & 1\\

\enddata
\tablenotetext{a}{All entries computed using linear values of DIB equivalent widths}
\tablenotetext{b}{Excluding $\rho$ Oph A, $\theta^1$ Ori C, and HD 37061}
\end{deluxetable}


\begin{deluxetable}{lcccccc}
\tablenum{6}
\tablecolumns{7}
\tablewidth{0pt}
\tablecaption{Correlation Data with $\lambda$5780.5\tablenotemark{a,b}}
\tablehead{
\colhead{DIB} &
\colhead{Correlation Coefficient} &
\colhead{Reduced $\chi^2$} &
\colhead{Number of Sight Lines} &
\colhead{Correlation Coefficient\tablenotemark{c}} &
\colhead{a\tablenotemark{d}} &
\colhead{b\tablenotemark{d}}
}
\startdata
5705.1 & 0.98  & 2.20 & 91 & 0.98 & $-0.25 \pm 1.07$ & $0.228 \pm 0.004$ \\
6204.5 & 0.97  &  6.92 & 119 & 0.97 & $-0.24 \pm 0.88$ & $0.412 \pm 0.004$ \\
6196.0 & 0.97  & 11.66 & 116 & 0.97 & $-0.63 \pm 0.17$ & $0.111 \pm 0.001$ \\
6613.6 & 0.96  & 28.05 & 119 & 0.96 & $-13.39 \pm 0.52$ & $0.467 \pm 0.003$ \\
6283.8 & 0.96  &   3.82 & 125 & 0.96 & $28.24 \pm 5.8$ & $2.32 \pm 0.03$ \\
5487.7 & 0.95  & 4.86 & 93 & 0.95 & $-2.29 \pm 1.13$ & $0.233 \pm 0.005$ \\
5797.1 & 0.93  & 26.35 & 113 & 0.93 & $-1.65 \pm 0.72$ & $0.384 \pm 0.003$ \\
\enddata
\tablenotetext{a}{All values computed using linear values of DIB equivalent widths}
\tablenotetext{b}{Data in columns 2, 3, 4, 6, and 7 exclude $\rho$ Oph A, $\theta^1$ Ori C, and HD 37061}
\tablenotetext{c}{Includes all sight lines}
\tablenotetext{d}{Coefficients for $W$(DIB) = a + b$\times W_{\lambda}(5780.5)$}
\end{deluxetable}


\end{document}